\pdfoutput=1
\documentclass[aps,prb,twocolumn,showpacs,superscriptaddress]{revtex4-1}

\newcommand{\half}{\frac{1}{2}}

\usepackage[pdftex]{graphicx}
\usepackage[pdftex]{epsfig}
\usepackage{epstopdf}
\usepackage{color}
\usepackage{amssymb,amsmath}
\usepackage{graphicx}
\usepackage{dcolumn}
\usepackage{multirow}
\usepackage{hyperref}

\begin{document}

\def\simlt{\mathrel{\lower .3ex \rlap{$\sim$}\raise .5ex \hbox{$<$}}}
\def\half{ \frac {1}{2} }
\title{Multivalley effective mass theory simulation of donors in silicon}

\author{John King Gamble}
\email{jkgambl@sandia.gov}
\affiliation{These authors contributed equally to this work.}
\affiliation{Center for Computing Research, Sandia National Laboratories, Albuquerque, NM 87185, USA}
\author{N. Tobias Jacobson}
\email{ntjacob@sandia.gov}
\affiliation{These authors contributed equally to this work.}
\affiliation{Center for Computing Research, Sandia National Laboratories, Albuquerque, NM 87185, USA}
\author{Erik Nielsen}
\affiliation{Sandia National Laboratories, Albuquerque, NM 87185, USA}
\author{Andrew D. Baczewski}
\affiliation{Center for Computing Research, Sandia National Laboratories, Albuquerque, NM 87185, USA}
\author{Jonathan E. Moussa}
\affiliation{Center for Computing Research, Sandia National Laboratories, Albuquerque, NM 87185, USA}
\author{In\`es Monta\~no}
\affiliation{Sandia National Laboratories, Albuquerque, NM 87185, USA}
\author{Richard P. Muller}
\affiliation{Center for Computing Research, Sandia National Laboratories, Albuquerque, NM 87185, USA}


\begin{abstract}
Last year, Salfi \emph{et al.} made the first direct measurements of a donor wave function and found extremely good theoretical agreement with atomistic tight-binding [Salfi \emph{et al.}, Nat. Mater. \textbf{13}, 605 (2014)]. 
Here, we show that multi-valley effective mass theory, applied properly, does achieve close agreement with tight-binding and hence gives reliable predictions.
To demonstrate this, we variationally solve the coupled six-valley Shindo-Nara equations, including silicon's full Bloch functions.
Surprisingly, we find that including the full Bloch functions necessitates a tetrahedral, rather than spherical, donor central cell correction to accurately reproduce the experimental energy spectrum of a phosphorus impurity in silicon.
We cross-validate this method against atomistic tight-binding calculations, showing that the two theories agree well for the calculation of donor-donor tunnel coupling.
Further, we benchmark our results by performing a statistical uncertainty analysis, confirming that derived quantities such as the wave function profile and tunnel couplings are robust with respect to variational energy fluctuations.
Finally, we apply this method to exhaustively enumerate the tunnel coupling for all donor-donor configurations within a large search volume,
demonstrating conclusively that the tunnel coupling has no spatially stable regions.
Though this instability is problematic for reliably coupling donor pairs for two-qubit operations,
we identify specific target locations where donor qubits can be placed with scanning tunneling microscopy technology to achieve reliably large tunnel couplings.
\end{abstract}

\maketitle

\section{Introduction}
Building on the pioneering work of Kohn and Luttinger,\cite{Kohn1955}
and later motivated by the promise of using donors in silicon for quantum information processing,\cite{Kane1998,Pla2012,Pla2013,Dehollain2014,Gonz2013b} 
researchers continue to develop and improve effective mass theories (EMTs).\cite{Ning1971,Pantelides1974,Shindo1976,Friesen2005,Wellard2005,Debernardi2006,Hui2013,Klymenko2014,Pica2014,Saraiva2014}
These theories are appealing both because they provide excellent physical intuition and because they require minimal computational resources to implement.  In most cases, parameters for the theory are chosen so that the donor binding energies match experimental values.\cite{Jagannath1981,Mayur1993}
Differences in the particular approximations adopted have led to dramatic discrepancies (\emph{e.g.}, orders of magnitude differences in exchange oscillations between Refs.~\onlinecite{Pica2014} and~\onlinecite{Wellard2005}), and one is often left wondering which, if any, of multiple, seemingly well-justified theories to believe.  It is this muddy picture, where small changes to the theories lead to large differences in outcomes, that has cast doubt upon whether EMT is well suited to making quantitative predictions.  

Past work has compared the wavefunction as predicted by both EMT and more sophisticated theories to experiment via the contact hyperfine interaction, which serves to probe the wavefunction directly at the donor site.
Early work by Feher\cite{Feher:1959} and later by Hale and Mieher\cite{Hale1969,Hale1969_2} compared the contact hyperfine predicted by Kohn-Luttinger EMT to electron nuclear double resonance experiments, obtaining rough qualitative agreement. 
Later, Ivey and Mieher used tight-binding (TB)\cite{Ivey1975,Ivey1975_2} to gain better agreement to the previous experimental data of Hale and Mieher\cite{Hale1969,Hale1969_2}.
More recently, Overhof and Uwe,\cite{Overhof:2004} Huebl \emph{et al.},\cite{Huebl:2006} and Assali \emph{et al.}\cite{Assali2011} studied the contact hyperfine interaction using \emph{ab initio} density functional theory,
resulting in much-improved experimental agreement.
Friesen\cite{Friesen2005} developed a multi-valley effective mass theory that was capable of studying the Stark shift of the contact hyperfine interaction.
Advancements in TB theory\cite{Martins2005,Manchero1999} enabled the detailed study by Rahman \emph{et al.}\cite{Rahman2007} of this Stark shift, which obtained excellent agreement with experiment. 
Finally, a more sophisticated EMT approach due to Pica \emph{et al.}\cite{Pica2014_2} also obtained experimental agreement for the contact hyperfine Stark shift.

However, full spatial wavefunctions have seldom been compared between theories, perhaps with the excuse that the results lacked strong experimental support.
The picture is different now: last year, Salfi \emph{et al.}~performed the first direct measurement of a donor wave function \cite{Salfi2014} and found excellent theoretical agreement with atomistic tight-binding simulation.\cite{klimeck2002development}  
Hence, it is important now to ask whether EMT can replicate the results of atomistic tight-binding;
the primary contention of this work is that it can when applied properly. 

By avoiding unjustified approximations, we present an effective mass framework that, in addition to matching experimental energies,\cite{Jagannath1981,Mayur1993} agrees well with atomistic tight-binding theory over the full spatial wavefunction.\cite{Salfi2014}
This agreement is of critical importance, since while operation of a single donor qubit requires a well-controlled hyperfine coupling, coupling two donor qubits depends upon reliable control over the electronic wave function far from the impurity site. 
The combined computational efficiency and accuracy of our EMT allows us to survey all possible donor-donor position combinations within a large search volume.
This is a critically important problem, since the coupling strength varies on the atomic scale due to silicon's six-fold conduction band valley degeneracy. 
Hence, for a given range of desired coupling strengths, our calculations allow for quantitative estimates of yield in the face of uncertain donor placement.

This paper is organized as follows. In Sec.~\ref{sec:SN}, we describe Shindo-Nara effective mass theory. 
First, Sec.~\ref{sec:SNbloch} details the calculation of silicon's Bloch function, and how they are included in the theory.
Here, we pay special attention to common approximations to the Bloch functions and where they lead to inconsistent results.
Sec.~\ref{sec:SNcc} discusses the role of the central cell correction in our calculation, and in particular our tetrahedrally symmetric variant necessary to reproduce the energy spectrum of phosphorus donors in silicon when the full Bloch functions of silicon are used.
Sec.~\ref{sec:SNvar} describes our variational solution to the theory,
including a statistical uncertainty quantification (UQ) procedure that demonstrates the stability of our results and a comparison to NEMO-3D atomistic tight-binding calculations.
Sec.~\ref{sec:tunnelCouplings} presents results of our calculations of donor-donor tunnel couplings. 
In Sec.~\ref{sec:tunnelCouplingsenum}, we first cross-validate our results using NEMO-3D calculations and check for stability using our UQ procedure.
We then detail the exhaustive enumeration of the tunnel coupling of all possible relative positions between a phosphorus donor at the origin
and a second donor at all lattice locations throughout a 30 nm surrounding cube of silicon.
After that, Sec.~\ref{sec:tunnelCouplingsenumstraggle} studies the implications of these results on the feasibility of achieving large donor-donor coupling when faced with uncertain donor placement.
Finally, in Sec.~\ref{sec:summary} we summarize our results and offer concluding remarks.

\section{Shindo-Nara effective mass theory} \label{sec:SN}

\begin{figure*}[tb]
\includegraphics[width= 1.0 \linewidth]{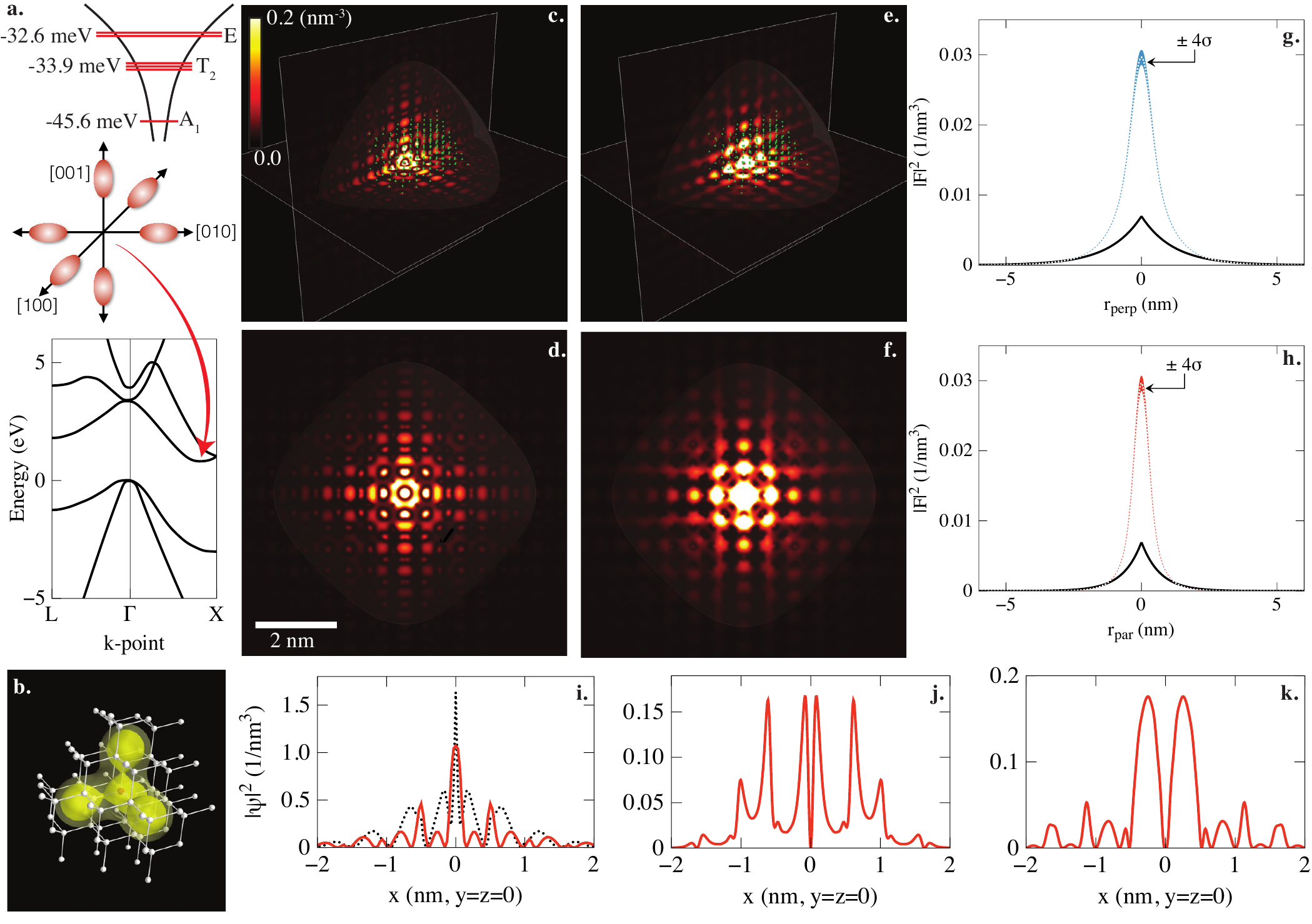}
\caption{\label{oneDonor}(Color online) 
Multi-valley effective mass calculations for a single phosphorus donor in silicon.
(a), Sketch of the band structure of silicon and the resulting donor physics. 
The conduction band valleys are initially six-fold degenerate; valley orbit coupling causes level splitting due to the sharp confinement of an impurity potential. 
The resulting energy levels for phosphorus are shown.
(b), Our converged donor potential, including the central cell correction, which exhibits tetrahedral symmetry. The constant energy surfaces shown are $-0.5$ (outer contour), $-1.0$ (middle contour), and $-4.0$~eV (central contour), respectively.
(c-d), Multi-valley effective mass ground state for a single phosphorus donor in silicon. (c) shows a side view, while (d) shows a top-down view of the $x-y$ plane. The silicon lattice is superposed toward the center of the plots for scale; the white curtain indicates when the envelope $|F|^2$ is one percent of its maximum value. 
(e-f), Atomistic tight-binding simulations corresponding to (c-d), performed in NEMO-3D and visualized using the atomic orbitals of Ref.~\onlinecite{Nielsen2012}. The envelope curtain is copied from (c-d) for comparison.
(g-h), Cuts along the parallel and perpendicular directions of the envelope function in one conduction band valley. The dashed lines are the effective mass theory from the present work; the shaded bands are $\pm 4 \sigma$ statistical uncertainty limits, determined by the UQ techniques described in Appendix~\ref{sec:crossValuq}. The lower bold curves show the corresponding Kohn-Luttinger envelope functions, for comparison.
(i-k), Cuts along the $x-$axis of the entire effective mass electron density for effective mass (solid curves) and NEMO-3D (dotted curve in i). 
(i) shows the $A_1$ ground state, (j) shows one of the three degenerate $T_2$ first excited states, and (k) shows one of the two degenerate $E$ first excited states.
}
\end{figure*}

\begin{figure*}[tb]
\includegraphics[width= 0.8 \linewidth]{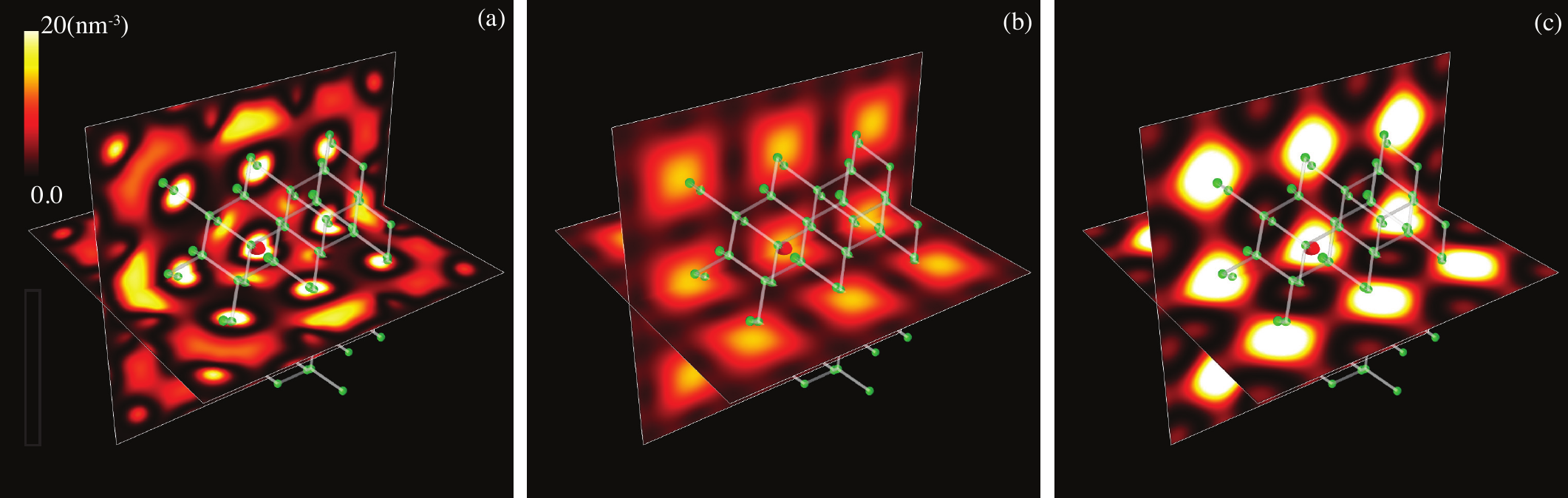}
\caption{\label{blochFunctions}
(Color online) The total Bloch function density, with the silicon lattice superimposed on the plots for scale.
Panel (a) shows well-converged Bloch functions, including high-frequency terms due to their periodic parts. 
Panel (b) truncates to \emph{form factors}, where each pair $u^*_{\mathbf k_0^l}(\mathbf r)u_{\mathbf k_0^j}(\mathbf r)$ is set equal to its constant-frequency component. Panel (c) simplifies the situation further, using \emph{trivial} ($u_{\mathbf k_0^j}(\mathbf r)=1$) Bloch functions.
As shown in panels (b) and (c), these represent drastic approximations, so it is not surprising that calculations using them yield results different from those using the Bloch function in panel (a).
}
\end{figure*}

The central tenet of effective mass theory for a low-energy conduction electron in silicon is that its wave function $\psi(\mathbf{r})$ has support only within the  vicinity of the six equivalent valley minima,\cite{Kohn1955} sketched in Fig.~\ref{oneDonor}(a):
\begin{equation}
\psi(\mathbf{r}) = \sum_{j=1}^6 F_j(\mathbf r) \phi_j(\mathbf r).
\end{equation}
Here, the sum runs over the six valley minima $\mathbf k_0^j$, located $0.84 \times 2 \pi/a$ along the cartesian axes ($a=0.543$~nm is the cubic unit cell length of Si), and $\phi_j(\mathbf r) =u_{\mathbf k_0^j}(\mathbf r) e^{i \mathbf k_0^j \cdot r}$ is the Bloch function belonging to the minimum of the $j$th valley. The prefactors $F_j(\mathbf r)$ are called \emph{envelope functions}, and are slowly varying on the length scale of the lattice. The multi-valley EMT formalism we use here was first derived by Shindo and Nara.\cite{Shindo1976}
The central equation of their theory is:
\begin{equation}\label{eq:SN}
E  F_l(\mathbf r)  = \left( \hat{\mathbf{T}}_l + U(\mathbf r ) \right) F_l(\mathbf r) + \sum_{j \in \pm \lbrace x, y, z \rbrace} V^{VO}_{lj}(\mathbf r) F_j(\mathbf r),
\end{equation}
which is an effective Schr\"odinger-like equation for the envelope functions. Here, $\hat{\mathbf{T}}_l$ is the kinetic energy operator of the $l$th valley, where for example $\hat{\mathbf{T}}_{+z} = -\frac{\hbar^{2}}{2m_{\perp}} \big( \frac{\partial^{2}}{\partial x^{2}} + \frac{\partial^{2}}{\partial y^{2}} + \gamma \frac{\partial^{2}}{\partial z^{2}} \big)$ with $\gamma = m_{\perp}/m_{\parallel}$ the ratio of effective masses, $U(\mathbf r)$ is the external (non-crystal) potential energy, and $V^{VO}(\mathbf{r})$ is the valley-orbit coupling 
$V^{VO}_{lj}(\mathbf{r}) = \phi_l^*(\mathbf r) \phi_j(\mathbf r)  U(\mathbf r)$.
To solve Eq.~(\ref{eq:SN}), we first need to compute $V^{VO}$, which requires the computation of the Bloch functions $\phi_j(\mathbf r)$ of silicon as well as the potential energy $U(\mathbf r)$. In the next section, we detail the calculation of $\phi_j(\mathbf r)$ within density functional theory. After that, we describe the calculation of $U(\mathbf r)$, including an appropriate central cell correction.

\subsection{Calculation and approximation of silicon Bloch functions}\label{sec:SNbloch}
We calculated the Bloch function at the conduction band minimum of pure bulk silicon using Kohn-Sham density functional theory 
within a variety of different approximations. 
We employed a plane wave basis in all cases, using both the Vienna Ab-initio Simulation Package  (VASP) \cite{Kresse1996,Kresse1999} and 
Quantum Espresso \cite{QE2009} packages to check for consistency among results. 
In VASP, we used the Projector Augmented Wave (PAW) formalism \cite{Blochl1994} to treat the electron-ion interaction, whereas in Quantum Espresso we used a variety of norm-conserving pseudopotentials 
(NCPP). 
While the plane wave coefficients are strictly $\ell^2$-normalized in NCPP calculations, they are not in PAW. 
Rather than explicitly including the effect of the PAW projectors on the normalization, we rescaled coefficients to 
achieve strict normalization, i.e. $1 \equiv \sum_\mathbf G \left| A^j_\mathbf G \right| ^2 $, with the Fourier components $ A^j_\mathbf G$ defined in Eq.~(\ref{blochFuncDef}).
 In a typical case, the uncorrected norm is within $3\%$ of unity due to the delocalized nature of the 
conduction band minimum orbital. In both methods, we performed calculations using different parameterizations of the local density approximation (LDA) and generalized gradient approximation (GGA) exchange-correlation functionals, as well as the hybrid Heyd-Scuseria-Ernzerhof (HSE) functional.\cite{Heyd2003}

For each pseudization scheme and functional, we used the following procedure for calculating the Bloch function. 
First, we performed an initial highly-converged self-consistent calculation to generate the Kohn-Sham potential which reproduces the ground state electronic density and energy 
of pure bulk silicon within a specific exchange-correlation approximation. 
Our criterion for self-consistency was that the change in the total energy between cycles be less than 1 $\mu$eV. 
We increased the plane wave cutoff and number of k-points in the first Brillouin zone until the total energy was converged to less than 1 meV/atom. 
Next, we performed a second, non-self-consistent calculation with the fixed Kohn-Sham Hamiltonian at 1000 equally-spaced k-points between the $\Gamma$ and X points, i.e., along $\Delta$. 
We then assessed the resultant Kohn-Sham orbital energies of the lowest conduction band to determine the location of the conduction 
band minimum. 
Finally, we extracted the Bloch function as the plane wave coefficients of this particular orbital.

The coefficients generated using different codes and functionals show a high degree of consistency. 
Applying a uniform phase shift to make the $(0,0,0)$ coefficient of each Bloch function real, the $\ell^2$ distance between any given pair of Bloch functions was 
$\approx$ 0.025 or less, representing approximately $2.5 \%$ relative error.
We observed a similar degree of consistency with results published elsewhere.\cite{Saraiva2011}
While the different exchange-correlation approximations utilized give a reasonably accurate description of the equilibrium lattice
constant of silicon as well as the ordering and character of its near-gap bands, they vary dramatically in precise value of the
band gap.\cite{Heyd2005} This deficiency seems to be irrelevant to the Bloch function of interest. 
Due to this consistency, there is no specific choice that appears to give a best Bloch function.
We choose to use the results of the PAW/HSE calculation, which are tabulated in a supplementary data file.

It is worth commenting on the use of Kohn-Sham orbitals in this capacity. 
Strictly speaking, the direct physical significance of 
these orbitals is limited, as they are solely intended to serve an auxiliary role in representing the interacting electronic 
density which is the real quantity of interest in DFT.\cite{Kohn1965} Here, we are simply using Kohn-Sham DFT as a convenient tool to generate 
an effective mean-field Hamiltonian, the eigenfunctions of which have qualitative atomically-resolved features that are needed by our 
effective mass theory. The high degree of consistency of relevant quantities between calculations gives us confidence that this approach is reasonable.

The contact hyperfine interaction between the spins of the electron and donor nucleus, nonzero for the $A_{1}$ ground state and zero for all other five states in the $1s$ manifold, is proportional to the charge density of the full electron wavefunction at the donor nucleus, $\vert \psi(0)\vert^{2}$. 
Past work in \emph{ab initio} DFT has found good agreement with experiment,\cite{Overhof:2004,Huebl:2006,Assali2011} demonstrating that $\vert \psi(0)\vert^{2}$ is well-understood.
Effective mass theory typically does not attempt to predict contact hyperfine, since doing so requires detailed knowledge of the Bloch function near the atomic core.
Rather, a \emph{bunching factor} is defined,\cite{Assali2011} which can either be tuned to experiment or calculated from DFT, and empirically accounts for a larger amplitude of the electronic wave function near the core than would be predicted from EMT alone.

Using a bunching factor facilitates comparison of contact hyperfine with experiment, which appears promising.\cite{Pica2014_2} 
We find that our calculated value of $\vert \psi_{A_{1}}(0)\vert^{2} = 1.0 \ \mathrm{nm^{-3}}$ is consistent with a Bloch function bunching factor at the donor of about $440$ in order to give the measured contact hyperfine interaction strength of 117.5 $\mathrm{MHz}$.\cite{Pica2014,Saraiva2014}
The precise value of this bunching factor is not physically meaningful, since the atomic core lies within the augmentation sphere (with radius 0.1054 nm) used in the computation of our Bloch functions. 
For this calculation, we include only the plane wave part of the full PAW wave function. 
In particular, we note that the dominant contribution of the central cell correction lies well outside the augmentation sphere around the donor, so we expect the central cell parameterization to not be significantly sensitive to the wave function form near the nucleus.
Since this work is concerned primarily with the electronic wave function away from the donor site, for simplicity we do not employ the more sophisticated techniques required to resolve the contact hyperfine coupling.

A common approximation performed in EMT treatments involves substantially simplifying the Bloch functions. 
However, such approximations lead to uncontrolled error, significantly disrupting the reliability of EMT.
To avoid this, we decompose the periodic part of the Bloch functions into plane wave components \cite{Saraiva2011}, which are computed as described above using density functional theory.
We discuss the practical cost of this procedure in more detail later in this section.

One of the most common approximations employed in effective mass theory is to simplify the form of the valley-orbit coupling by using only an approximation for the  Bloch function product $ \phi_l^*(\mathbf r) \phi_j(\mathbf r) $. 
Here, each of the Bloch functions is specified by\cite{Saraiva2011}
\begin{align}\label{blochFuncDef}
\phi_j(\mathbf r) &= u_{\mathbf k_0^j}(\mathbf r) e^{i \mathbf k_0^j \cdot r} \nonumber \\
& = e^{i \mathbf k_0^j \cdot r}  \sum_{\mathbf G} A_{\mathbf G}^j e^{i \mathbf G \cdot r},
\end{align}
where $\mathbf G$ is a reciprocal lattice vector. 
We compute the set of Fourier coefficients $\{  A_{\mathbf G}^j \}$that determine $u_{\mathbf k_0^j}(\mathbf r)$ using density functional theory (as described above), and we list our coefficients in a supplementary data file. 

The most drastic way to approximate the Bloch function product is to set $\phi_j(\mathbf r) \approx e^{i \mathbf k_0^j \cdot r}$, amounting to \emph{trivial} Bloch functions. 
Another approximation is to write the product as $\phi_l^*(\mathbf r) \phi_j(\mathbf r)  \approx  C_{lj} e^{i (\mathbf k_0^j-\mathbf k_0^l) \cdot r}$, where the factors $C_{lj}$ are called \emph{form factors}. 
In Fig.~\ref{blochFunctions}a, we plot the full Bloch function density $\sum_j \left| \phi_j(\mathbf r) \right|^2$. In Figs.~\ref{blochFunctions}b-c we plot the same quantity with the form factor and trivial approximations, respectively. As can be seen from these plots, the Bloch functions under either approximation are not qualitatively similar to the full Bloch function. In addition, 
using these approximate Bloch functions with a central cell correction tuned for more converged Bloch functions results in substantial energy discrepancies, with the ground state energy approximately 5 meV too positive when using trivial Bloch functions and approximately 10 meV too positive when using form factor Bloch functions. Although re-converging a central cell correction with these approximate Bloch functions can improve this discrepancy, it remains the case that these approximations are neither well-justified nor well-controlled.

We next address the practical cost of including the full Bloch functions in our theory. 
This addition does not change the dimensionality of the Hamiltonian being diagonalized, and the only additional computational cost is associated with the evaluation of the valley-orbit matrix elements. 
In practice, when we compute matrix elements, we truncate the series
\begin{equation}
\phi_l^*(\mathbf r) \phi_j(\mathbf r)  =  \sum_{\mathbf G, \mathbf G'} \left(A_{\mathbf G'}^l \right)^* A_{\mathbf G}^j e^{i (\mathbf k_0^j-\mathbf k_0^l+\mathbf G - \mathbf G' ) \cdot \mathbf r}
\end{equation}
to include all terms up to $|\mathbf G - \mathbf G' | \leq 4.4 \times 2 \pi / a$, which we find to be well-converged. 
By grouping elements by the $\mathbf G - \mathbf G'$ vectors, this results in about 100 terms, and hence
 the evaluation of a matrix element is $\sim100$ times slower than it would be for trivial Bloch functions. 
Even so, the total cost is still negligible relative to atomistic methods in which the dimensionality of the Hamiltonian scales with the total number of valence orbitals comprising a supercell of the sample (in this case, millions), whereas the dimensionality 
of our Hamiltonian scales with the total number of donor basis sets included in this same volume in EMT (in this case, two sets). 
It should also be emphasized that the calculation of the Bloch functions is strictly restricted to precomputation and tabulation.
The associated DFT calculations do not contribute to the computational cost of our method in practice.

\subsection{Calculation of the central cell correction}\label{sec:SNcc}
The attractive binding potential $U(\mathbf r)$ due to a donor in silicon is well approximated at long distances as a bulk-screened Coulomb potential. 
However, close to the impurity the dielectric screening effect of silicon is lessened, and the potential is enhanced. 
The deviation of the potential $U(\mathbf r)$ from bulk-screened Coulomb form at short distances is called a \emph{central cell correction}.\cite{Ning1971,Pantelides1974}
In order to reproduce experimentally observed donor energy levels,\cite{Jagannath1981,Mayur1993} we tune the central cell correction using a nested variational optimization. 
It is worth noting that central cell corrections tuned with the crude approximations of the Bloch functions outlined above do not maintain experimental agreement when the full, correct Bloch functions are used. 
Likewise, for a central cell correction tuned to the full Bloch functions, using the approximate forms results in markedly different energies.
To date, all EMT studies of electron donor energy levels that employ a central cell correction have assumed a spherically-symmetric or contact ($\delta$-function) correction.\cite{Ning1971,Pantelides1974,Friesen2005,Hui2013,Pica2014} 
However, to accurately reproduce experimentally observed donor binding energies to within experimental measurement uncertainties, we find it necessary to allow for a tetrahedrally-symmetric central cell correction  (Fig.~\ref{oneDonor}(b)), as anticipated in Refs.~\onlinecite{Castner2009, Greenman2013}.

\begin{table}[tb]
\caption{\label{ccparams}Parameters for central cell correction $U_{\mathrm{cc}}(\mathbf{r})$}
\begin{tabular}{|r|l|}
\hline
$A_{0}$ & -1.2837 \ meV \\
$A_{1}$ & -2642.0 \ meV \\
$a$ & 0.12857 \ nm \\
$b$ & 0.21163 \ nm \\
$c$ & 0.09467 \ nm \\
\hline
\end{tabular}
\end{table}

We determined the central cell correction for the phosphorus donor by the following nested variational procedure:
Inner optimization: Given a central cell correction, construct the total potential, solve the full coupled effective mass equation variationally using a Gaussian basis with 6 $1s$-type orbitals and one $2s$-type orbital.
Outer optimization: Vary the form of the central cell correction in order to optimize the experimental energies for phosphorus shown in Fig.~1(a).

Far from the donor, the donor's binding potential takes the form of a bulk-screened Coulomb potential, $U_{\mathrm{c}}(r) = -e^{2} / (4 \pi \epsilon_{\mathrm{Si}} r)$, where $\epsilon_{\mathrm{Si}} = 11.7 \epsilon_0$ is silicon's dielectric constant, $\epsilon_0$ is the permittivity of free space, and $e$ is the electron's charge. Near the donor, the local potential deviates from this simple $1/r$ behavior, as a result of reduced dielectric screening from the silicon lattice and complex reorganization of the local electronic structure.\cite{Pantelides1974,Greenman2013} 

To describe this effect, we include a central cell correction $U_{\mathrm{cc}}(\mathbf{r})$, such that the full donor impurity potential takes the form $U(\mathbf{r}) = U_{c}(r) + U_{\mathrm{cc}}(\mathbf{r})$. Due to the tetrahedral symmetry of the covalent bonding between the donor and the neighboring silicon atoms in the lattice, we allow for the central cell correction $U_{\mathrm{cc}}(\mathbf{r})$ to be tetrahedrally symmetric, to be contrasted with the more restrictive spherical symmetry assumed in previous studies.\cite{Ning1971,Pantelides1974,Hui2013,Pica2014}
We find that this tetrahedral symmetry is necessary in order to obtain the correct donor binding energies when the full Bloch function is considered. 
Specifically, unlike with trivial Bloch functions, we find that the donor valley splitting cannot be made large enough to match experiment using a spherically symmetric central cell.

We allow the central cell correction to be a function of five parameters,
\begin{equation}\label{eq:Central_cell_correction}
U_{\mathrm{cc}}(\mathbf{r}) = A_{0} e^{-r^{2}/(2a^{2})} + A_{1} \sum_{i=1}^{4} e^{-\vert \mathbf{r} - b \mathbf{t}_{i} \vert^{2} / (2c^{2})},
\end{equation}
where $\mathbf{t}_{i} \in \lbrace (1,1,1), (-1,1,-1), (1,-1,-1),(-1,-1,1) \rbrace$. 
This potential takes the form of a Gaussian centered at the origin plus four identical Gaussians centered at points along the bond directions.  
We choose this Gaussian basis for the central cell correction as a convenient means of representing a smooth potential with compact support.
In our convention for the lattice coordinates, we take the position of the sites of the primitive unit cell to be $(0,0,0)$ and $(a/4)(1,1,1)$, where $a=0.543 \ \mathrm{nm}$. The tetrahedral directions $\mathbf{t}_{i}$ are taken to be oriented along the bonds, for the donor assumed to be located at the coordinate $(0,0,0)$. If the donor is located at a site equivalent to the coordinate $(a/4)(1,1,1)$, the tetrahedral directions must be inverted, $\mathbf{t}_{i} \to -\mathbf{t}_{i}$, to preserve agreement with the bond directions. 

Following the nested optimization process described earlier, we list the parameters for the tetrahedrally-symmetric central cell correction  of Eq. (\ref{eq:Central_cell_correction}) in Table~\ref{ccparams}. 
Note in particular that the strength of the tetrahedral lobes, $A_{1}$, is large compared to the central spherical term $A_{0}$.
This underscores the importance of allowing our central cell to have tetrahedral symmetry. The nested variational approach we used to determine the central cell parameters is underdetermined, as we use 5 unknowns to satisfy 3 constraints. Hence, we began the optimization with physically reasonable initial parameters, and terminated the optimization when the donor energies were well within experimental uncertainties. To confirm that our solution is stable, we developed a statistical UQ technique (Appendix \ref{sec:crossValuq}), which we use throughout this study.

We remark now on an inconsistency inherent to using $\delta$-function contact potentials to fit the energy levels, as in Refs \onlinecite{Fritzsche1962, Friesen2005}. In three dimensions, it is well known that attractive potentials of the form $U(r) = -\alpha \delta^{(3)}(r)$ exhibit infinitely many bound states, with a ground state of infinitely negative energy. While this approach captures the essential physics necessary for first-order perturbation theory, it is inconsistent with any sufficiently rich variational optimization for the orbital basis.

\subsection{Variational solution}\label{sec:SNvar}

\begin{figure*}[tb]
\includegraphics[width= 1.0 \linewidth]{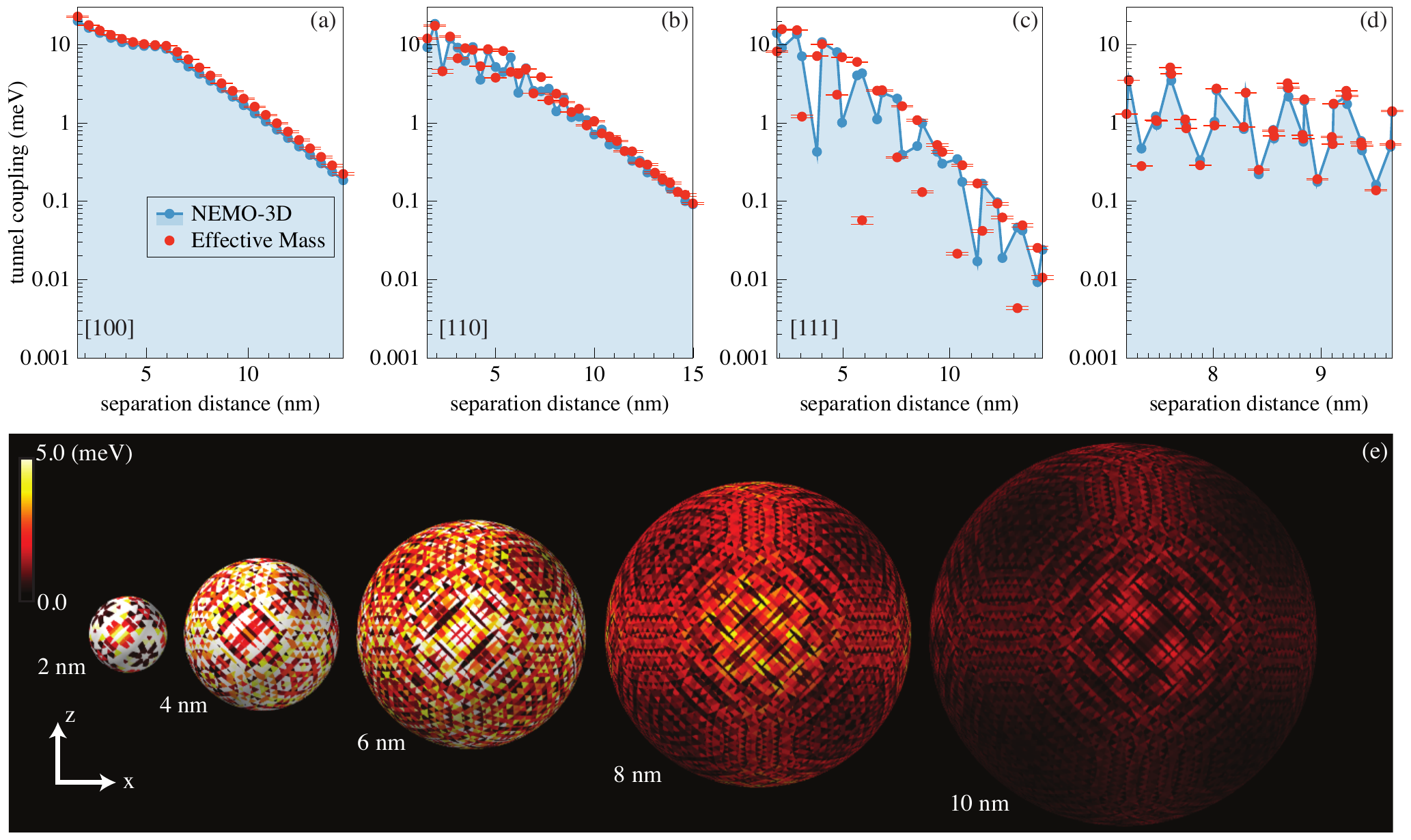}
\caption{\label{twoDonors}(Color online) 
Tunnel couplings computed for two phosphorus donors in silicon. (a-d), Comparison of tunnel couplings computed within multi-valley effective mass theory (points with error bars) and NEMO-3D atomistic tight-binding (connected points with no error bars). Here, the tunnel coupling is defined as the energy difference between the first excited state and ground state of the one-electron, two-donor problem. Panel (a) shows tunnel coupling along the [100] direction, panel (b) along [110], and panel (c) along [111]. Panel (d) depicts typical random instances, not along any particular direction. In all cases, the atomistic and effective mass theory exhibits very similar trends and magnitude of oscillations. Along [111], there appears to be a phase discrepancy, likely due to differing placements of the conduction band minima (see the main text for details). The error bars on the effective mass predictions are $\pm 4 \sigma$ statistical uncertainty limits, determined by the UQ techniques described in Appendix~\ref{sec:crossValuq}. (e), Exhaustive tunnel coupling enumeration for two phosphorus donors. Here, we placed one donor at the origin and the second at every possible point within a 30~nm cube surrounding it ($\sim 1.3$ million instances). The spherical shells show cuts (with nearest-neighbor interpolation) of the tunnel coupling at fixed donor separation distances. The tunnel coupling is highly oscillatory, and there is no large region of stability in the tunnel coupling. The full results of the enumeration are tabulated in a supplementary data file.
}
\end{figure*}

\begin{figure*}[tb]
\includegraphics[width= 1.0 \linewidth]{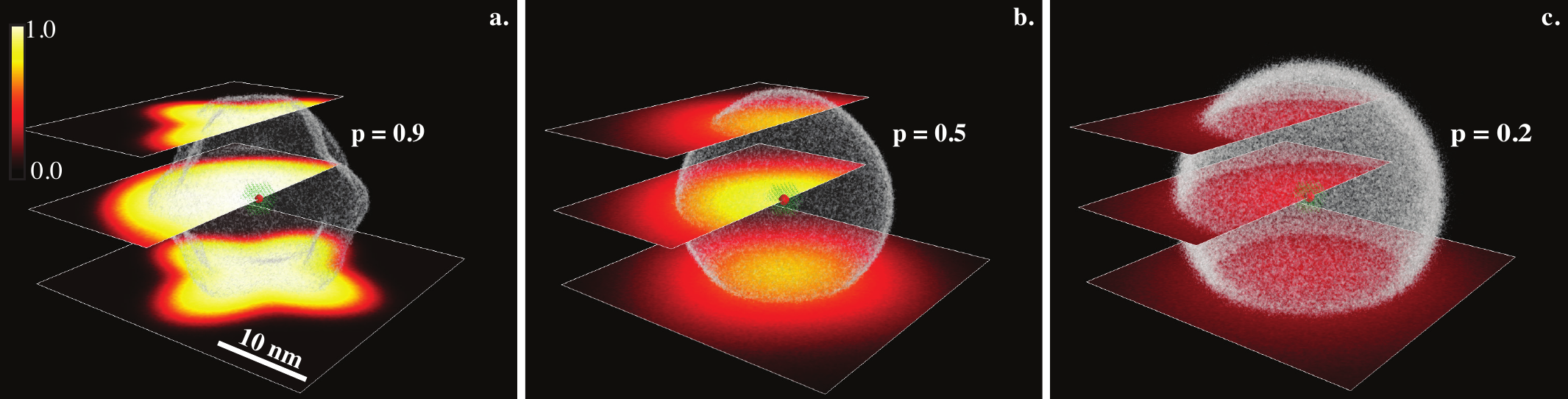}
\caption{\label{tunnelTarget}(Color online)
Probability of achieving large tunnel coupling with uncertain donor placement.
 In each panel, one phosphorus donor is placed at the origin and a second is placed at lattice sites within a surrounding 30~nm cube. 
The placement of the second donor is uncertain. 
The plotted probability is that of obtaining $t>0.1$~meV for a Gaussian distribution of donor placements as a function of the distribution's center. 
The lower bound of $0.1$~meV is chosen to be about an order of magnitude larger than typical dilution refrigerator electron temperatures.
We performed 20000-shot Monte-Carlo, sampling from a 3D isotropic Gaussian distribution with varying widths: panel (a) corresponds to 1~nm, panel (b) to 5~nm, and panel (c) to 10~nm straggle.
Panel (a) depicts an experimentally realistic straggle for STM-based donor placement, while panels (b) and (c) depict the results of increasing donor straggle.
The white curtain shown in each plot indicates the contour of constant probability as labeled. 
These results show that STM placement can ensure large tunnel coupling with high yield, while ion implantation technology can only ever achieve low yield, rendering ion implantation ineffective for deterministic use.
}
\end{figure*}

\begin{table}[tb]
\caption{\label{functionParams}Parameters for the variational Cartesian Gaussian envelope basis}
\begin{tabular}{|c|c|c|c|}
\hline
Index & $ (n_{x},n_{y},n_{z})$ & $\alpha_{\perp} \ \mathrm{(nm^{-2})}$ & $\alpha_{\parallel}  \ \mathrm{(nm^{-2})}$ \\
\hline
1 & $(0,0,0)$ & 3.48877 & 6.93542 \\
2 & $(0,0,0)$ & 0.84055 & 3.06020 \\
3 & $(0,0,0)$ & 0.39326 & 1.23742 \\
4 & $(0,0,0)$ & 0.03096 & 0.12142 \\
5 & $(0,0,0)$ & 0.01209 & 0.06195 \\
6 & $(0,0,0)$ & 0.00732 & 0.03747 \\
7 & $(2,0,0)$ & 0.20364 & 0.70775 \\
8 & $(0,2,0)$ & 0.20364 & 0.70775 \\
9 & $(0,0,2)$ & 0.20364 & 0.70775 \\
\hline
\end{tabular}
\end{table}

Now that we have computed both the Bloch functions and the central cell correction, we are equipped to 
solve Eq.~(\ref{eq:SN}).
We do so variationally, by expanding each $F_j$ over a finite orbital basis set of size $N$,
\begin{equation}
 F_j(\mathbf r) = \sum_{\nu=1}^{N} A_{(j,\nu)} F_{(j,\nu)}(\mathbf r),
\end{equation}
where the coefficients $ A_{(j,\nu)}$ are unknowns to be determined. 
For each phosphorus atom and valley, we construct a basis from nine atom-centered Cartesian Gaussian functions.
For an atom at the origin and the $+z$ valley, for example, we have 
\begin{equation}
F_{(+z,\nu)}(\mathbf{r}) = \mathcal{N}  x^{n_{x}} y^{n_{y}} z^{n_{z}} e^{-\alpha_{\perp} (x^{2} + y^{2})}e^{-\alpha_{\parallel} z^{2}},
\end{equation}
where the normalization factor $\mathcal{N}$ is chosen such that $\int_{\mathrm{all \ space}} d^{3}r \ \vert F_{(j,\nu)} \vert^{2} = 1$.
By symmetry, the orbital basis for one valley is equivalent up to a coordinate permutation to that of other valleys. 

Within this basis, we express Eq.~\ref{eq:SN} as the generalized eigenvalue problem
\begin{equation}
\sum_{j,\nu} \mathbf{H}_{(l,\mu),(j,\nu)} A_{(j,\nu)} = E \sum_{j,\nu}  \mathbf{S}_{(l,\mu),(j,\nu)} A_{(j,\nu)},
\end{equation}
where the Hamiltonian matrix elements are 
\begin{align}
 \mathbf{H}_{(l,\mu),(j,\nu)} &= \int d^3 r F^*_{(l,\mu)}(\mathbf r)  F_{(j,\nu)}(\mathbf r)  \\
 &\times \left[ \left( \hat{\mathbf{T}}_l + U(\mathbf r ) \right)\delta_{l,j} + V^{VO}_{lj}(\mathbf r) \right] \nonumber
\end{align}
and the overlap matrix, block-diagonal with respect to the valleys, is given by
\begin{equation}
\mathbf{S}_{(l,\mu),(j,\nu)} =  \int d^3 r F^*_{(l,\mu)}(\mathbf r)  F_{(j,\nu)}(\mathbf r) \delta_{l,j}.
\end{equation}

Using this matrix formalism, for a fixed $U(\mathbf r)$, we perform a nonlinear optimization to minimize the ground state energy with respect to the nonlinear basis parameters (the $\alpha_{\perp}$ and $\alpha_{\parallel}$ parameters above).
For each step in the nonlinear optimization we solve the linear matrix problem. 
Hence, for any basis ansatz we determine the optimal linear combination of basis functions that minimizes the ground state energy.
The linear combinations of basis functions detailed in Table~\ref{functionParams}  that form the lowest six energy eigenstates are tabulated in a supplementary data file.

Figs.~\ref{oneDonor}(c-d) illustrate the charge density of the ground ($A_1$) state of a phosphorus donor in silicon given by our calculations. 
For comparison, we solved the same problem using the atomistic tight-binding code NEMO-3D,\cite{klimeck2002development}
as shown in Figs.~\ref{oneDonor}(e-f) and detailed in Appendix~\ref{sec:crossValNemo}; 
we find visual agreement between the two very different methods.  
In Figs~\ref{oneDonor}(g-h), we show the variation of the envelope function along the principal axes of one of the six identical envelope functions of the ground state. 
For comparison, we also plot the envelope functions of Kohn and Luttinger \cite{Kohn1955}, with decay constants found in Ref.~\onlinecite{Baena2012}.
The error bars shown are determined by a Monte-Carlo UQ procedure detailed in Appendix~\ref{sec:crossValuq}.
As has been anticipated,\cite{Saraiva2014} the states are more strongly peaked than those of Kohn and Luttinger, but are more weakly peaked than other recent calculations that assume approximate Bloch functions and a spherically-symmetric central cell correction.\cite{Pica2014}
Figs.~\ref{oneDonor}(i-k) show variation along the $x$-axis of the charge density of the ground ($A_{1}$) state, one of the three degenerate first excited ($T_{2}$) states, and one of the two degenerate higher excited ($E$) states, respectively. 

\section{Study of donor-donor tunnel couplings}\label{sec:tunnelCouplings}

\subsection{Exhaustive enumeration of tunnel couplings}\label{sec:tunnelCouplingsenum}

Next, we compute the tunnel coupling $t$ between two phosphorus donors using the multi-valley EMT framework. 
We define tunnel coupling as the energy difference between the one-electron first excited and ground states of two donors. 
Earlier work predicted significant sensitivity with respect to donor placement of tunnel coupling \cite{Hu2005} as well as exchange,\cite{Koiller2001} and our results for the tunnel coupling confirm this. 
Tunnel coupling and exchange are correlated through their mutual dependence on the strength of overlap between states localized to each donor.\cite{Li2010} 
Figs.~\ref{twoDonors}a-d compare with the results of NEMO-3D; we plot the tunnel coupling along three high-symmetry directions, and in addition a sampling of random instances at various separation distances. Agreement is quantitatively very strong, with the exception of Fig.~\ref{twoDonors}c. 
There, the results appear to be out of phase, although the magnitude of oscillation and trend are very similar. 
Of special note is that both theories agree perfectly on where the transition between the strong- and weak-coupling regimes occurs, in which the first excited state changes character.\cite{Klymenko2014,Saraiva2014}
Shown as a kink in the curves of Fig~\ref{twoDonors}a, this transition occurs at about 6 nm separation along [100].

Having cross-validated EMT predictions for tunnel coupling, we next leverage the computational efficiency of our EMT to perform an exhaustive enumeration of tunnel couplings within a specified volume.
In Fig.~\ref{twoDonors}e, we position one donor at the origin and sweep the second through all valid locations located in an enclosing 30 nm cube, resulting in $\sim 1.3$ million donor placements. 
To visualize these data, we plot the tunnel coupling on concentric shells of varying radii using nearest-neighbor interpolation. 
For quantum computing applications, since donor placement has experimental uncertainty (placement straggle), it is desirable for tunnel coupling to be stable under small perturbations of position.
Unfortunately, we see here that the tunnel couplings are highly oscillatory. 
Using this exhaustive analysis, we conclude that there does not exist a sizable region of adjacent donor placements that exhibits stability with respect to straggle, an issue that we will explore in more detail in the next section.

\subsection{Statistical analysis of placement straggle}\label{sec:tunnelCouplingsenumstraggle}

Since two-qubit gates rely on large couplings between donors \cite{Kane1998}, the preceding calculations cast severe doubt on their experimental viability.
Having ruled out deterministically stable tunnel couplings, we turn now to statistical analysis.
We accept a donor placement configuration if the tunnel coupling satisfies $t>0.1$~meV, which is roughly an order of magnitude larger than typical dilution refrigerator electron temperatures.
We then quantify the probability of obtaining this range given a target donor location and straggle.
Straggle is determined in practice by the technology used for donor placement.
For scanning tunneling microscope (STM) placement, a conservative overestimate of the straggle is $\sim 1$~nm.\cite{Oberbeck2004}
In contrast to this precision placement, ion implantation techniques typically have spreads of tens of nm.\cite{Bielejec2010}

To study the effects of different placement technologies on achieving high tunnel couplings, in Fig.~\ref{tunnelTarget} we show the probability of achieving $t>0.1$~meV for three different donor straggles: 1~nm in panel (a), 5~nm in panel (b), and 10~nm in panel (c).
In each case, the straggle distribution is taken to be an isotropic Gaussian distribution.
We determine the probabilities shown by dividing our 30 nm placement cube into a $201 \! \times \! 201\! \times \! 201$ grid of target donor locations and perform 20000 Monte Carlo samples of the tunnel coupling at each point.
For STM-compatible placement we find large regions where acceptably large tunnel coupling occurs with high probability, while for the typical placement uncertainty of ion implantation we do not.  
We therefore expect that achieving $t>0.1$ meV is practical using STM placement but impractical using ion implantation.

\section{Summary}\label{sec:summary}

We have demonstrated that properly parameterized effective mass theory obtains results that agree quantitatively with both experimental energy spectroscopy and atomistic tight-binding theory \cite{klimeck2002development} that has been recently validated against direct measurement.\cite{Salfi2014}
After benchmarking against tight-binding, we leveraged the computational efficiency of EMT to exhaustively enumerate about 1.3 million donor placements, a task not presently feasible with atomistic methods.
We show that although there do not exist any regions of stable tunnel coupling, there do exist regions where experimentally realistic donor placement uncertainty results in large tunnel couplings with high yield.
By means of a reliable, physically transparent, and high-throughput statistical survey, this work illustrates that effective mass theory is well suited to quantitative explorations of donor physics that are impractical to solve using more computationally intensive techniques.

\section*{Acknowledgements}
We thank A.~Saraiva, W.~Witzel, S.~Coppersmith, M.~Friesen, M.~Carroll, A.~Frees, T.~Boykin, J.~Aidun, and P.~Schultz for useful discussions and comments on the manuscript,
and R.~Rahman and G.~Klimeck for assistance and support with the NEMO-3D simulations.
The simulations presented in this work were performed, in part, on Sandia National Laboratories' Red Sky computing cluster.
This work was supported, in part, by the Laboratory Directed Research and Development program at Sandia National Laboratories.  Sandia National Laboratories is a multi-program laboratory managed and operated by Sandia Corporation, a wholly owned subsidiary of Lockheed Martin Corporation, for the U.S. Department of Energy's National Nuclear Security Administration under contract DE-AC04-94AL85000.  
 
 \appendix

\section{Details of the NEMO-3D atomistic tight binding calculations}\label{sec:crossValNemo}

We performed the tight-binding calculations using the Nanoelectronic Modeling Tool (NEMO-3D).\cite{klimeck2002development} For the work presented here we utilized a set of 10 localized orbital bases $sp^3d^5s^*$  on a 3D relaxed silicon (diamond structure) atomistic lattice.\cite{Boykin2004}  The phosphorus donor is modeled by a Coulomb potential screened by the dielectric constant of Si and with a cutoff potential $U_0$ at the donor site.\cite{Rahman2011} 

\section{Details of the statistical benchmarking and uncertainty quantification (UQ)}\label{sec:crossValuq}

The nested variational optimization we use to determine the central cell correction and wavefunctions takes as input only the experimentally measured energy levels. To quantify the degree to which fitting to the energies constrains the wavefunction, we performed an uncertainty quantification analysis for the wavefunctions induced by the experimental error bars for the energies.

Taking a conservative estimate of the experimental error bars synthesized from the literature,\cite{Jagannath1981,Mayur1993} we assume that the energy levels $( A_{1}, T_{2}, E)$ take the form of a multivariate normal distribution with mean $\mu =(-45.59, -33.89, -32.58 ) \ \mathrm{meV}$ and covariances $\mathbb{E}\left[ (E_{i} - \mu_{i})^{2}\right] = \Delta^{2}$ and $\mathbb{E}\left[ \big( (E_{i} - \mu_{i}) - (E_{j} - \mu_{j}) \big)^{2}\right]_{i \neq j} = \delta^{2}$, with $\delta = 0.05 \ \mathrm{meV}$ and $\Delta = 0.2 \ \mathrm{meV}$. Here, $\Delta$ and $\delta$ quantify the uncertainty in the absolute value of the energy levels and energy differences, respectively. Given this distribution for the energy levels, we randomly perturb all central cell correction parameters by 1\% and evaluate the energy spectrum and eigenstates. This variation of the central cell correction parameters is chosen to adequately sample over the support of the energy level distribution. We then associate the probability density of the energy levels with the envelope function and use this weighted ensemble of envelope functions to determine qualitative error bars for the wavefunctions.

\bibliography{EMTPaper.bib}

\begin{thebibliography}{52}%
\makeatletter
\providecommand \@ifxundefined [1]{%
 \@ifx{#1\undefined}
}%
\providecommand \@ifnum [1]{%
 \ifnum #1\expandafter \@firstoftwo
 \else \expandafter \@secondoftwo
 \fi
}%
\providecommand \@ifx [1]{%
 \ifx #1\expandafter \@firstoftwo
 \else \expandafter \@secondoftwo
 \fi
}%
\providecommand \natexlab [1]{#1}%
\providecommand \enquote  [1]{``#1''}%
\providecommand \bibnamefont  [1]{#1}%
\providecommand \bibfnamefont [1]{#1}%
\providecommand \citenamefont [1]{#1}%
\providecommand \href@noop [0]{\@secondoftwo}%
\providecommand \href [0]{\begingroup \@sanitize@url \@href}%
\providecommand \@href[1]{\@@startlink{#1}\@@href}%
\providecommand \@@href[1]{\endgroup#1\@@endlink}%
\providecommand \@sanitize@url [0]{\catcode `\\12\catcode `\$12\catcode
  `\&12\catcode `\#12\catcode `\^12\catcode `\_12\catcode `\%12\relax}%
\providecommand \@@startlink[1]{}%
\providecommand \@@endlink[0]{}%
\providecommand \url  [0]{\begingroup\@sanitize@url \@url }%
\providecommand \@url [1]{\endgroup\@href {#1}{\urlprefix }}%
\providecommand \urlprefix  [0]{URL }%
\providecommand \Eprint [0]{\href }%
\providecommand \doibase [0]{http://dx.doi.org/}%
\providecommand \selectlanguage [0]{\@gobble}%
\providecommand \bibinfo  [0]{\@secondoftwo}%
\providecommand \bibfield  [0]{\@secondoftwo}%
\providecommand \translation [1]{[#1]}%
\providecommand \BibitemOpen [0]{}%
\providecommand \bibitemStop [0]{}%
\providecommand \bibitemNoStop [0]{.\EOS\space}%
\providecommand \EOS [0]{\spacefactor3000\relax}%
\providecommand \BibitemShut  [1]{\csname bibitem#1\endcsname}%
\let\auto@bib@innerbib\@empty
\bibitem [{\citenamefont {Kohn}\ and\ \citenamefont
  {Luttinger}(1955)}]{Kohn1955}%
  \BibitemOpen
  \bibfield  {author} {\bibinfo {author} {\bibfnamefont {W.}~\bibnamefont
  {Kohn}}\ and\ \bibinfo {author} {\bibfnamefont {J.}~\bibnamefont
  {Luttinger}},\ }\href {http://prola.aps.org/abstract/PR/v98/i4/p915\_1}
  {\bibfield  {journal} {\bibinfo  {journal} {Phys. Rev.}\ }\textbf {\bibinfo
  {volume} {98}} (\bibinfo {year} {1955})}\BibitemShut {NoStop}%
\bibitem [{\citenamefont {Kane}(1998)}]{Kane1998}%
  \BibitemOpen
  \bibfield  {author} {\bibinfo {author} {\bibfnamefont {B.~E.}\ \bibnamefont
  {Kane}},\ }\href {\doibase 10.1038/30156} {\bibfield  {journal} {\bibinfo
  {journal} {Nature}\ }\textbf {\bibinfo {volume} {393}},\ \bibinfo {pages}
  {133} (\bibinfo {year} {1998})}\BibitemShut {NoStop}%
\bibitem [{\citenamefont {Pla}\ \emph {et~al.}(2012)\citenamefont {Pla},
  \citenamefont {Tan}, \citenamefont {Dehollain}, \citenamefont {Lim},
  \citenamefont {Morton}, \citenamefont {Jamieson}, \citenamefont {Dzurak},\
  and\ \citenamefont {Morello}}]{Pla2012}%
  \BibitemOpen
  \bibfield  {author} {\bibinfo {author} {\bibfnamefont {J.~J.}\ \bibnamefont
  {Pla}}, \bibinfo {author} {\bibfnamefont {K.~Y.}\ \bibnamefont {Tan}},
  \bibinfo {author} {\bibfnamefont {J.~P.}\ \bibnamefont {Dehollain}}, \bibinfo
  {author} {\bibfnamefont {W.~H.}\ \bibnamefont {Lim}}, \bibinfo {author}
  {\bibfnamefont {J.~J.~L.}\ \bibnamefont {Morton}}, \bibinfo {author}
  {\bibfnamefont {D.~N.}\ \bibnamefont {Jamieson}}, \bibinfo {author}
  {\bibfnamefont {A.~S.}\ \bibnamefont {Dzurak}}, \ and\ \bibinfo {author}
  {\bibfnamefont {A.}~\bibnamefont {Morello}},\ }\href {\doibase
  10.1038/nature11449} {\bibfield  {journal} {\bibinfo  {journal} {Nature}\
  }\textbf {\bibinfo {volume} {489}},\ \bibinfo {pages} {541} (\bibinfo {year}
  {2012})}\BibitemShut {NoStop}%
\bibitem [{\citenamefont {Pla}\ \emph {et~al.}(2013)\citenamefont {Pla},
  \citenamefont {Tan}, \citenamefont {Dehollain}, \citenamefont {Lim},
  \citenamefont {Morton}, \citenamefont {Zwanenburg}, \citenamefont {Jamieson},
  \citenamefont {Dzurak},\ and\ \citenamefont {Morello}}]{Pla2013}%
  \BibitemOpen
  \bibfield  {author} {\bibinfo {author} {\bibfnamefont {J.~J.}\ \bibnamefont
  {Pla}}, \bibinfo {author} {\bibfnamefont {K.~Y.}\ \bibnamefont {Tan}},
  \bibinfo {author} {\bibfnamefont {J.~P.}\ \bibnamefont {Dehollain}}, \bibinfo
  {author} {\bibfnamefont {W.~H.}\ \bibnamefont {Lim}}, \bibinfo {author}
  {\bibfnamefont {J.~J.~L.}\ \bibnamefont {Morton}}, \bibinfo {author}
  {\bibfnamefont {F.~a.}\ \bibnamefont {Zwanenburg}}, \bibinfo {author}
  {\bibfnamefont {D.~N.}\ \bibnamefont {Jamieson}}, \bibinfo {author}
  {\bibfnamefont {A.~S.}\ \bibnamefont {Dzurak}}, \ and\ \bibinfo {author}
  {\bibfnamefont {A.}~\bibnamefont {Morello}},\ }\href {\doibase
  10.1038/nature12011} {\bibfield  {journal} {\bibinfo  {journal} {Nature}\
  }\textbf {\bibinfo {volume} {496}},\ \bibinfo {pages} {334} (\bibinfo {year}
  {2013})}\BibitemShut {NoStop}%
\bibitem [{\citenamefont {Dehollain}\ \emph {et~al.}(2014)\citenamefont
  {Dehollain}, \citenamefont {Muhonen}, \citenamefont {Tan}, \citenamefont
  {Saraiva}, \citenamefont {Jamieson}, \citenamefont {Dzurak},\ and\
  \citenamefont {Morello}}]{Dehollain2014}%
  \BibitemOpen
  \bibfield  {author} {\bibinfo {author} {\bibfnamefont {J.~P.}\ \bibnamefont
  {Dehollain}}, \bibinfo {author} {\bibfnamefont {J.~T.}\ \bibnamefont
  {Muhonen}}, \bibinfo {author} {\bibfnamefont {K.~Y.}\ \bibnamefont {Tan}},
  \bibinfo {author} {\bibfnamefont {A.}~\bibnamefont {Saraiva}}, \bibinfo
  {author} {\bibfnamefont {D.~N.}\ \bibnamefont {Jamieson}}, \bibinfo {author}
  {\bibfnamefont {A.~S.}\ \bibnamefont {Dzurak}}, \ and\ \bibinfo {author}
  {\bibfnamefont {A.}~\bibnamefont {Morello}},\ }\href {\doibase
  10.1103/PhysRevLett.112.236801} {\bibfield  {journal} {\bibinfo  {journal}
  {Phys. Rev. Lett.}\ }\textbf {\bibinfo {volume} {112}},\ \bibinfo {pages}
  {236801} (\bibinfo {year} {2014})}\BibitemShut {NoStop}%
\bibitem [{\citenamefont {Gonzalez-Zalba}\ \emph {et~al.}(2013)\citenamefont
  {Gonzalez-Zalba}, \citenamefont {Saraiva}, \citenamefont {Heiss},
  \citenamefont {Calder\'{o}n}, \citenamefont {Koiller},\ and\ \citenamefont
  {Ferguson}}]{Gonz2013b}%
  \BibitemOpen
  \bibfield  {author} {\bibinfo {author} {\bibfnamefont {M.~F.}\ \bibnamefont
  {Gonzalez-Zalba}}, \bibinfo {author} {\bibfnamefont {A.}~\bibnamefont
  {Saraiva}}, \bibinfo {author} {\bibfnamefont {D.}~\bibnamefont {Heiss}},
  \bibinfo {author} {\bibfnamefont {M.~J.}\ \bibnamefont {Calder\'{o}n}},
  \bibinfo {author} {\bibfnamefont {B.}~\bibnamefont {Koiller}}, \ and\
  \bibinfo {author} {\bibfnamefont {A.~J.}\ \bibnamefont {Ferguson}},\ }\href
  {http://arxiv.org/abs/1312.4589} {\ ,\ \bibinfo {pages} {9} (\bibinfo {year}
  {2013})},\ \Eprint {http://arxiv.org/abs/1312.4589} {arXiv:1312.4589}
  \BibitemShut {NoStop}%
\bibitem [{\citenamefont {Ning}\ and\ \citenamefont {Sah}(1971)}]{Ning1971}%
  \BibitemOpen
  \bibfield  {author} {\bibinfo {author} {\bibfnamefont {T.}~\bibnamefont
  {Ning}}\ and\ \bibinfo {author} {\bibfnamefont {C.}~\bibnamefont {Sah}},\
  }\href {\doibase 10.1103/PhysRevB.4.3468} {\bibfield  {journal} {\bibinfo
  {journal} {Phys. Rev. B}\ }\textbf {\bibinfo {volume} {4}},\ \bibinfo {pages}
  {3468} (\bibinfo {year} {1971})}\BibitemShut {NoStop}%
\bibitem [{\citenamefont {Pantelides}\ and\ \citenamefont
  {Sah}(1974)}]{Pantelides1974}%
  \BibitemOpen
  \bibfield  {author} {\bibinfo {author} {\bibfnamefont {S.}~\bibnamefont
  {Pantelides}}\ and\ \bibinfo {author} {\bibfnamefont {C.}~\bibnamefont
  {Sah}},\ }\href {\doibase 10.1103/PhysRevB.10.621} {\bibfield  {journal}
  {\bibinfo  {journal} {Phys. Rev. B}\ }\textbf {\bibinfo {volume} {10}},\
  \bibinfo {pages} {621} (\bibinfo {year} {1974})}\BibitemShut {NoStop}%
\bibitem [{\citenamefont {Shindo}\ and\ \citenamefont
  {Nara}(1976)}]{Shindo1976}%
  \BibitemOpen
  \bibfield  {author} {\bibinfo {author} {\bibfnamefont {K.}~\bibnamefont
  {Shindo}}\ and\ \bibinfo {author} {\bibfnamefont {H.}~\bibnamefont {Nara}},\
  }\href {\doibase 10.1143/JPSJ.40.1640} {\bibfield  {journal} {\bibinfo
  {journal} {J. Phys. Soc. Japan}\ }\textbf {\bibinfo {volume} {40}},\ \bibinfo
  {pages} {1640} (\bibinfo {year} {1976})}\BibitemShut {NoStop}%
\bibitem [{\citenamefont {Friesen}(2005)}]{Friesen2005}%
  \BibitemOpen
  \bibfield  {author} {\bibinfo {author} {\bibfnamefont {M.}~\bibnamefont
  {Friesen}},\ }\href {\doibase 10.1103/PhysRevLett.94.186403} {\bibfield
  {journal} {\bibinfo  {journal} {Phys. Rev. Lett.}\ }\textbf {\bibinfo
  {volume} {94}},\ \bibinfo {pages} {186403} (\bibinfo {year}
  {2005})}\BibitemShut {NoStop}%
\bibitem [{\citenamefont {Wellard}\ and\ \citenamefont
  {Hollenberg}(2005)}]{Wellard2005}%
  \BibitemOpen
  \bibfield  {author} {\bibinfo {author} {\bibfnamefont {C.~J.}\ \bibnamefont
  {Wellard}}\ and\ \bibinfo {author} {\bibfnamefont {L.~C.~L.}\ \bibnamefont
  {Hollenberg}},\ }\href {\doibase 10.1103/PhysRevB.72.085202} {\bibfield
  {journal} {\bibinfo  {journal} {Phys. Rev. B}\ }\textbf {\bibinfo {volume}
  {72}},\ \bibinfo {pages} {085202} (\bibinfo {year} {2005})}\BibitemShut
  {NoStop}%
\bibitem [{\citenamefont {Debernardi}\ \emph {et~al.}(2006)\citenamefont
  {Debernardi}, \citenamefont {Baldereschi},\ and\ \citenamefont
  {Fanciulli}}]{Debernardi2006}%
  \BibitemOpen
  \bibfield  {author} {\bibinfo {author} {\bibfnamefont {A.}~\bibnamefont
  {Debernardi}}, \bibinfo {author} {\bibfnamefont {A.}~\bibnamefont
  {Baldereschi}}, \ and\ \bibinfo {author} {\bibfnamefont {M.}~\bibnamefont
  {Fanciulli}},\ }\href {\doibase 10.1103/PhysRevB.74.035202} {\bibfield
  {journal} {\bibinfo  {journal} {Phys. Rev. B}\ }\textbf {\bibinfo {volume}
  {74}},\ \bibinfo {pages} {035202} (\bibinfo {year} {2006})}\BibitemShut
  {NoStop}%
\bibitem [{\citenamefont {Hui}(2013)}]{Hui2013}%
  \BibitemOpen
  \bibfield  {author} {\bibinfo {author} {\bibfnamefont {H.}~\bibnamefont
  {Hui}},\ }\href {\doibase 10.1016/j.ssc.2012.10.023} {\bibfield  {journal}
  {\bibinfo  {journal} {Solid State Commun.}\ }\textbf {\bibinfo {volume}
  {154}},\ \bibinfo {pages} {19} (\bibinfo {year} {2013})}\BibitemShut
  {NoStop}%
\bibitem [{\citenamefont {Klymenko}\ and\ \citenamefont
  {Remacle}(2014)}]{Klymenko2014}%
  \BibitemOpen
  \bibfield  {author} {\bibinfo {author} {\bibfnamefont {M.~V.}\ \bibnamefont
  {Klymenko}}\ and\ \bibinfo {author} {\bibfnamefont {F.}~\bibnamefont
  {Remacle}},\ }\href {\doibase 10.1088/0953-8984/26/6/065302} {\bibfield
  {journal} {\bibinfo  {journal} {J. Phys. Condens. Matter}\ }\textbf {\bibinfo
  {volume} {26}},\ \bibinfo {pages} {065302} (\bibinfo {year}
  {2014})}\BibitemShut {NoStop}%
\bibitem [{\citenamefont {Pica}\ \emph
  {et~al.}(2014{\natexlab{a}})\citenamefont {Pica}, \citenamefont {Lovett},
  \citenamefont {Bhatt},\ and\ \citenamefont {Lyon}}]{Pica2014}%
  \BibitemOpen
  \bibfield  {author} {\bibinfo {author} {\bibfnamefont {G.}~\bibnamefont
  {Pica}}, \bibinfo {author} {\bibfnamefont {B.~W.}\ \bibnamefont {Lovett}},
  \bibinfo {author} {\bibfnamefont {R.~N.}\ \bibnamefont {Bhatt}}, \ and\
  \bibinfo {author} {\bibfnamefont {S.~A.}\ \bibnamefont {Lyon}},\ }\href
  {\doibase 10.1103/PhysRevB.89.235306} {\bibfield  {journal} {\bibinfo
  {journal} {Phys. Rev. B}\ }\textbf {\bibinfo {volume} {89}},\ \bibinfo
  {pages} {235306} (\bibinfo {year} {2014}{\natexlab{a}})}\BibitemShut
  {NoStop}%
\bibitem [{\citenamefont {Saraiva}\ \emph {et~al.}(2014)\citenamefont
  {Saraiva}, \citenamefont {Baena}, \citenamefont {Calder\'{o}n},\ and\
  \citenamefont {Koiller}}]{Saraiva2014}%
  \BibitemOpen
  \bibfield  {author} {\bibinfo {author} {\bibfnamefont {A.~L.}\ \bibnamefont
  {Saraiva}}, \bibinfo {author} {\bibfnamefont {A.}~\bibnamefont {Baena}},
  \bibinfo {author} {\bibfnamefont {M.~J.}\ \bibnamefont {Calder\'{o}n}}, \
  and\ \bibinfo {author} {\bibfnamefont {B.}~\bibnamefont {Koiller}},\ }\href
  {http://arxiv.org/abs/1407.8224} {\bibfield  {journal} {\bibinfo  {journal}
  {arXiv}\ }\textbf {\bibinfo {volume} {2}},\ \bibinfo {pages} {13} (\bibinfo
  {year} {2014})},\ \Eprint {http://arxiv.org/abs/1407.8224} {arXiv:1407.8224}
  \BibitemShut {NoStop}%
\bibitem [{\citenamefont {Jagannath}\ \emph {et~al.}(1981)\citenamefont
  {Jagannath}, \citenamefont {Grabowski},\ and\ \citenamefont
  {Ramdas}}]{Jagannath1981}%
  \BibitemOpen
  \bibfield  {author} {\bibinfo {author} {\bibfnamefont {C.}~\bibnamefont
  {Jagannath}}, \bibinfo {author} {\bibfnamefont {Z.~W.}\ \bibnamefont
  {Grabowski}}, \ and\ \bibinfo {author} {\bibfnamefont {A.~K.}\ \bibnamefont
  {Ramdas}},\ }\href {\doibase 10.1103/PhysRevB.23.2082} {\bibfield  {journal}
  {\bibinfo  {journal} {Phys. Rev. B}\ }\textbf {\bibinfo {volume} {23}},\
  \bibinfo {pages} {2082} (\bibinfo {year} {1981})}\BibitemShut {NoStop}%
\bibitem [{\citenamefont {Mayur}\ \emph {et~al.}(1993)\citenamefont {Mayur},
  \citenamefont {Sciacca}, \citenamefont {Ramdas},\ and\ \citenamefont
  {Rodriguez}}]{Mayur1993}%
  \BibitemOpen
  \bibfield  {author} {\bibinfo {author} {\bibfnamefont {A.~J.}\ \bibnamefont
  {Mayur}}, \bibinfo {author} {\bibfnamefont {M.~D.}\ \bibnamefont {Sciacca}},
  \bibinfo {author} {\bibfnamefont {A.~K.}\ \bibnamefont {Ramdas}}, \ and\
  \bibinfo {author} {\bibfnamefont {S.}~\bibnamefont {Rodriguez}},\ }\href
  {\doibase 10.1103/PhysRevB.48.10893} {\bibfield  {journal} {\bibinfo
  {journal} {Phys. Rev. B}\ }\textbf {\bibinfo {volume} {48}},\ \bibinfo
  {pages} {10893} (\bibinfo {year} {1993})}\BibitemShut {NoStop}%
\bibitem [{\citenamefont {Feher}(1959)}]{Feher:1959}%
  \BibitemOpen
  \bibfield  {author} {\bibinfo {author} {\bibfnamefont {G.}~\bibnamefont
  {Feher}},\ }\href {\doibase 10.1103/PhysRev.114.1219} {\bibfield  {journal}
  {\bibinfo  {journal} {Phys. Rev.}\ }\textbf {\bibinfo {volume} {114}},\
  \bibinfo {pages} {1219} (\bibinfo {year} {1959})}\BibitemShut {NoStop}%
\bibitem [{\citenamefont {Hale}\ and\ \citenamefont
  {Mieher}(1969{\natexlab{a}})}]{Hale1969}%
  \BibitemOpen
  \bibfield  {author} {\bibinfo {author} {\bibfnamefont {E.~B.}\ \bibnamefont
  {Hale}}\ and\ \bibinfo {author} {\bibfnamefont {R.~L.}\ \bibnamefont
  {Mieher}},\ }\href {\doibase 10.1103/PhysRev.184.739} {\bibfield  {journal}
  {\bibinfo  {journal} {Phys. Rev.}\ }\textbf {\bibinfo {volume} {184}},\
  \bibinfo {pages} {739} (\bibinfo {year} {1969}{\natexlab{a}})}\BibitemShut
  {NoStop}%
\bibitem [{\citenamefont {Hale}\ and\ \citenamefont
  {Mieher}(1969{\natexlab{b}})}]{Hale1969_2}%
  \BibitemOpen
  \bibfield  {author} {\bibinfo {author} {\bibfnamefont {E.~B.}\ \bibnamefont
  {Hale}}\ and\ \bibinfo {author} {\bibfnamefont {R.~L.}\ \bibnamefont
  {Mieher}},\ }\href {\doibase 10.1103/PhysRev.184.751} {\bibfield  {journal}
  {\bibinfo  {journal} {Phys. Rev.}\ }\textbf {\bibinfo {volume} {184}},\
  \bibinfo {pages} {751} (\bibinfo {year} {1969}{\natexlab{b}})}\BibitemShut
  {NoStop}%
\bibitem [{\citenamefont {Ivey}\ and\ \citenamefont
  {Mieher}(1975{\natexlab{a}})}]{Ivey1975}%
  \BibitemOpen
  \bibfield  {author} {\bibinfo {author} {\bibfnamefont {J.~L.}\ \bibnamefont
  {Ivey}}\ and\ \bibinfo {author} {\bibfnamefont {R.~L.}\ \bibnamefont
  {Mieher}},\ }\href {\doibase 10.1103/PhysRevB.11.822} {\bibfield  {journal}
  {\bibinfo  {journal} {Phys. Rev. B}\ }\textbf {\bibinfo {volume} {11}},\
  \bibinfo {pages} {822} (\bibinfo {year} {1975}{\natexlab{a}})}\BibitemShut
  {NoStop}%
\bibitem [{\citenamefont {Ivey}\ and\ \citenamefont
  {Mieher}(1975{\natexlab{b}})}]{Ivey1975_2}%
  \BibitemOpen
  \bibfield  {author} {\bibinfo {author} {\bibfnamefont {J.~L.}\ \bibnamefont
  {Ivey}}\ and\ \bibinfo {author} {\bibfnamefont {R.~L.}\ \bibnamefont
  {Mieher}},\ }\href {\doibase 10.1103/PhysRevB.11.849} {\bibfield  {journal}
  {\bibinfo  {journal} {Phys. Rev. B}\ }\textbf {\bibinfo {volume} {11}},\
  \bibinfo {pages} {849} (\bibinfo {year} {1975}{\natexlab{b}})}\BibitemShut
  {NoStop}%
\bibitem [{\citenamefont {Overhof}\ and\ \citenamefont
  {Gerstmann}(2004)}]{Overhof:2004}%
  \BibitemOpen
  \bibfield  {author} {\bibinfo {author} {\bibfnamefont {H.}~\bibnamefont
  {Overhof}}\ and\ \bibinfo {author} {\bibfnamefont {U.}~\bibnamefont
  {Gerstmann}},\ }\href {\doibase 10.1103/PhysRevLett.92.087602} {\bibfield
  {journal} {\bibinfo  {journal} {Phys. Rev. Lett.}\ }\textbf {\bibinfo
  {volume} {92}},\ \bibinfo {pages} {087602} (\bibinfo {year}
  {2004})}\BibitemShut {NoStop}%
\bibitem [{\citenamefont {Huebl}\ \emph {et~al.}(2006)\citenamefont {Huebl},
  \citenamefont {Stegner}, \citenamefont {Stutzmann}, \citenamefont {Brandt},
  \citenamefont {Vogg}, \citenamefont {Bensch}, \citenamefont {Rauls},\ and\
  \citenamefont {Gerstmann}}]{Huebl:2006}%
  \BibitemOpen
  \bibfield  {author} {\bibinfo {author} {\bibfnamefont {H.}~\bibnamefont
  {Huebl}}, \bibinfo {author} {\bibfnamefont {A.~R.}\ \bibnamefont {Stegner}},
  \bibinfo {author} {\bibfnamefont {M.}~\bibnamefont {Stutzmann}}, \bibinfo
  {author} {\bibfnamefont {M.~S.}\ \bibnamefont {Brandt}}, \bibinfo {author}
  {\bibfnamefont {G.}~\bibnamefont {Vogg}}, \bibinfo {author} {\bibfnamefont
  {F.}~\bibnamefont {Bensch}}, \bibinfo {author} {\bibfnamefont
  {E.}~\bibnamefont {Rauls}}, \ and\ \bibinfo {author} {\bibfnamefont
  {U.}~\bibnamefont {Gerstmann}},\ }\href {\doibase
  10.1103/PhysRevLett.97.166402} {\bibfield  {journal} {\bibinfo  {journal}
  {Phys. Rev. Lett.}\ }\textbf {\bibinfo {volume} {97}},\ \bibinfo {pages}
  {166402} (\bibinfo {year} {2006})}\BibitemShut {NoStop}%
\bibitem [{\citenamefont {Assali}\ \emph {et~al.}(2011)\citenamefont {Assali},
  \citenamefont {Petrilli}, \citenamefont {Capaz}, \citenamefont {Koiller},
  \citenamefont {Hu},\ and\ \citenamefont {{Das Sarma}}}]{Assali2011}%
  \BibitemOpen
  \bibfield  {author} {\bibinfo {author} {\bibfnamefont {L.~V.~C.}\
  \bibnamefont {Assali}}, \bibinfo {author} {\bibfnamefont {H.~M.}\
  \bibnamefont {Petrilli}}, \bibinfo {author} {\bibfnamefont {R.~B.}\
  \bibnamefont {Capaz}}, \bibinfo {author} {\bibfnamefont {B.}~\bibnamefont
  {Koiller}}, \bibinfo {author} {\bibfnamefont {X.}~\bibnamefont {Hu}}, \ and\
  \bibinfo {author} {\bibfnamefont {S.}~\bibnamefont {{Das Sarma}}},\ }\href
  {\doibase 10.1103/PhysRevB.83.165301} {\bibfield  {journal} {\bibinfo
  {journal} {Phys. Rev. B}\ }\textbf {\bibinfo {volume} {83}},\ \bibinfo
  {pages} {165301} (\bibinfo {year} {2011})}\BibitemShut {NoStop}%
\bibitem [{\citenamefont {Martins}\ \emph {et~al.}(2005)\citenamefont
  {Martins}, \citenamefont {Boykin}, \citenamefont {Klimeck},\ and\
  \citenamefont {Koiller}}]{Martins2005}%
  \BibitemOpen
  \bibfield  {author} {\bibinfo {author} {\bibfnamefont {A.~S.}\ \bibnamefont
  {Martins}}, \bibinfo {author} {\bibfnamefont {T.~B.}\ \bibnamefont {Boykin}},
  \bibinfo {author} {\bibfnamefont {G.}~\bibnamefont {Klimeck}}, \ and\
  \bibinfo {author} {\bibfnamefont {B.}~\bibnamefont {Koiller}},\ }\href
  {\doibase 10.1103/PhysRevB.72.193204} {\bibfield  {journal} {\bibinfo
  {journal} {Phys. Rev. B}\ }\textbf {\bibinfo {volume} {72}},\ \bibinfo
  {pages} {193204} (\bibinfo {year} {2005})}\BibitemShut {NoStop}%
\bibitem [{\citenamefont {Menchero}\ \emph {et~al.}(1999)\citenamefont
  {Menchero}, \citenamefont {Capaz}, \citenamefont {Koiller},\ and\
  \citenamefont {Chacham}}]{Manchero1999}%
  \BibitemOpen
  \bibfield  {author} {\bibinfo {author} {\bibfnamefont {J.~G.}\ \bibnamefont
  {Menchero}}, \bibinfo {author} {\bibfnamefont {R.~B.}\ \bibnamefont {Capaz}},
  \bibinfo {author} {\bibfnamefont {B.}~\bibnamefont {Koiller}}, \ and\
  \bibinfo {author} {\bibfnamefont {H.}~\bibnamefont {Chacham}},\ }\href
  {\doibase 10.1103/PhysRevB.59.2722} {\bibfield  {journal} {\bibinfo
  {journal} {Phys. Rev. B}\ }\textbf {\bibinfo {volume} {59}},\ \bibinfo
  {pages} {2722} (\bibinfo {year} {1999})}\BibitemShut {NoStop}%
\bibitem [{\citenamefont {Rahman}\ \emph {et~al.}(2007)\citenamefont {Rahman},
  \citenamefont {Wellard}, \citenamefont {Bradbury}, \citenamefont {Prada},
  \citenamefont {Cole}, \citenamefont {Klimeck},\ and\ \citenamefont
  {Hollenberg}}]{Rahman2007}%
  \BibitemOpen
  \bibfield  {author} {\bibinfo {author} {\bibfnamefont {R.}~\bibnamefont
  {Rahman}}, \bibinfo {author} {\bibfnamefont {C.~J.}\ \bibnamefont {Wellard}},
  \bibinfo {author} {\bibfnamefont {F.~R.}\ \bibnamefont {Bradbury}}, \bibinfo
  {author} {\bibfnamefont {M.}~\bibnamefont {Prada}}, \bibinfo {author}
  {\bibfnamefont {J.~H.}\ \bibnamefont {Cole}}, \bibinfo {author}
  {\bibfnamefont {G.}~\bibnamefont {Klimeck}}, \ and\ \bibinfo {author}
  {\bibfnamefont {L.~C.~L.}\ \bibnamefont {Hollenberg}},\ }\href {\doibase
  10.1103/PhysRevLett.99.036403} {\bibfield  {journal} {\bibinfo  {journal}
  {Phys. Rev. Lett.}\ }\textbf {\bibinfo {volume} {99}},\ \bibinfo {pages}
  {036403} (\bibinfo {year} {2007})}\BibitemShut {NoStop}%
\bibitem [{\citenamefont {Pica}\ \emph
  {et~al.}(2014{\natexlab{b}})\citenamefont {Pica}, \citenamefont {Wolfowicz},
  \citenamefont {Urdampilleta}, \citenamefont {Thewalt}, \citenamefont
  {Riemann}, \citenamefont {Abrosimov}, \citenamefont {Becker}, \citenamefont
  {Pohl}, \citenamefont {Morton}, \citenamefont {Bhatt}, \citenamefont {Lyon},\
  and\ \citenamefont {Lovett}}]{Pica2014_2}%
  \BibitemOpen
  \bibfield  {author} {\bibinfo {author} {\bibfnamefont {G.}~\bibnamefont
  {Pica}}, \bibinfo {author} {\bibfnamefont {G.}~\bibnamefont {Wolfowicz}},
  \bibinfo {author} {\bibfnamefont {M.}~\bibnamefont {Urdampilleta}}, \bibinfo
  {author} {\bibfnamefont {M.~L.~W.}\ \bibnamefont {Thewalt}}, \bibinfo
  {author} {\bibfnamefont {H.}~\bibnamefont {Riemann}}, \bibinfo {author}
  {\bibfnamefont {N.~V.}\ \bibnamefont {Abrosimov}}, \bibinfo {author}
  {\bibfnamefont {P.}~\bibnamefont {Becker}}, \bibinfo {author} {\bibfnamefont
  {H.-J.}\ \bibnamefont {Pohl}}, \bibinfo {author} {\bibfnamefont {J.~J.~L.}\
  \bibnamefont {Morton}}, \bibinfo {author} {\bibfnamefont {R.~N.}\
  \bibnamefont {Bhatt}}, \bibinfo {author} {\bibfnamefont {S.~A.}\ \bibnamefont
  {Lyon}}, \ and\ \bibinfo {author} {\bibfnamefont {B.~W.}\ \bibnamefont
  {Lovett}},\ }\href {\doibase 10.1103/PhysRevB.90.195204} {\bibfield
  {journal} {\bibinfo  {journal} {Phys. Rev. B}\ }\textbf {\bibinfo {volume}
  {90}},\ \bibinfo {pages} {195204} (\bibinfo {year}
  {2014}{\natexlab{b}})}\BibitemShut {NoStop}%
\bibitem [{\citenamefont {Salfi}\ \emph {et~al.}(2014)\citenamefont {Salfi},
  \citenamefont {Mol}, \citenamefont {Rahman}, \citenamefont {Klimeck},
  \citenamefont {Simmons}, \citenamefont {Hollenberg},\ and\ \citenamefont
  {Rogge}}]{Salfi2014}%
  \BibitemOpen
  \bibfield  {author} {\bibinfo {author} {\bibfnamefont {J.}~\bibnamefont
  {Salfi}}, \bibinfo {author} {\bibfnamefont {J.~A.}\ \bibnamefont {Mol}},
  \bibinfo {author} {\bibfnamefont {R.}~\bibnamefont {Rahman}}, \bibinfo
  {author} {\bibfnamefont {G.}~\bibnamefont {Klimeck}}, \bibinfo {author}
  {\bibfnamefont {M.~Y.}\ \bibnamefont {Simmons}}, \bibinfo {author}
  {\bibfnamefont {L.~C.~L.}\ \bibnamefont {Hollenberg}}, \ and\ \bibinfo
  {author} {\bibfnamefont {S.}~\bibnamefont {Rogge}},\ }\href {\doibase
  10.1038/nmat3941} {\bibfield  {journal} {\bibinfo  {journal} {Nat. Mater.}\
  }\textbf {\bibinfo {volume} {13}},\ \bibinfo {pages} {605} (\bibinfo {year}
  {2014})}\BibitemShut {NoStop}%
\bibitem [{\citenamefont {Klimeck}\ \emph {et~al.}(2002)\citenamefont
  {Klimeck}, \citenamefont {Oyafuso}, \citenamefont {Boykin}, \citenamefont
  {Bowen},\ and\ \citenamefont {von Allmen}}]{klimeck2002development}%
  \BibitemOpen
  \bibfield  {author} {\bibinfo {author} {\bibfnamefont {G.}~\bibnamefont
  {Klimeck}}, \bibinfo {author} {\bibfnamefont {F.}~\bibnamefont {Oyafuso}},
  \bibinfo {author} {\bibfnamefont {T.~B.}\ \bibnamefont {Boykin}}, \bibinfo
  {author} {\bibfnamefont {R.~C.}\ \bibnamefont {Bowen}}, \ and\ \bibinfo
  {author} {\bibfnamefont {P.}~\bibnamefont {von Allmen}},\ }\href@noop {}
  {\bibfield  {journal} {\bibinfo  {journal} {C. Comput. Model. Eng. Sci.}\
  }\textbf {\bibinfo {volume} {3}},\ \bibinfo {pages} {601} (\bibinfo {year}
  {2002})}\BibitemShut {NoStop}%
\bibitem [{\citenamefont {Nielsen}\ \emph {et~al.}(2012)\citenamefont
  {Nielsen}, \citenamefont {Rahman},\ and\ \citenamefont
  {Muller}}]{Nielsen2012}%
  \BibitemOpen
  \bibfield  {author} {\bibinfo {author} {\bibfnamefont {E.}~\bibnamefont
  {Nielsen}}, \bibinfo {author} {\bibfnamefont {R.}~\bibnamefont {Rahman}}, \
  and\ \bibinfo {author} {\bibfnamefont {R.~P.}\ \bibnamefont {Muller}},\
  }\href {\doibase http://dx.doi.org/10.1063/1.4759256} {\bibfield  {journal}
  {\bibinfo  {journal} {J. Appl. Phys.}\ }\textbf {\bibinfo {volume} {112}},\
  \bibinfo {eid} {114304} (\bibinfo {year} {2012})}\BibitemShut {NoStop}%
\bibitem [{\citenamefont {Kresse}\ and\ \citenamefont
  {Furthm{\"u}ller}(1996)}]{Kresse1996}%
  \BibitemOpen
  \bibfield  {author} {\bibinfo {author} {\bibfnamefont {G.}~\bibnamefont
  {Kresse}}\ and\ \bibinfo {author} {\bibfnamefont {J.}~\bibnamefont
  {Furthm{\"u}ller}},\ }\href {http://dx.doi.org/10.1103/PhysRevB.54.11169}
  {\bibfield  {journal} {\bibinfo  {journal} {Physical Review B}\ }\textbf
  {\bibinfo {volume} {54}},\ \bibinfo {pages} {11169} (\bibinfo {year}
  {1996})}\BibitemShut {NoStop}%
\bibitem [{\citenamefont {Kresse}\ and\ \citenamefont
  {Joubert}(1999)}]{Kresse1999}%
  \BibitemOpen
  \bibfield  {author} {\bibinfo {author} {\bibfnamefont {G.}~\bibnamefont
  {Kresse}}\ and\ \bibinfo {author} {\bibfnamefont {D.}~\bibnamefont
  {Joubert}},\ }\href {http://dx.doi.org/10.1103/PhysRevB.59.1758} {\bibfield
  {journal} {\bibinfo  {journal} {Physical Review B}\ }\textbf {\bibinfo
  {volume} {59}},\ \bibinfo {pages} {1758} (\bibinfo {year}
  {1999})}\BibitemShut {NoStop}%
\bibitem [{\citenamefont {Giannozzi}\ \emph {et~al.}(2009)\citenamefont
  {Giannozzi}, \citenamefont {Baroni}, \citenamefont {Bonini}, \citenamefont
  {Calandra}, \citenamefont {Car}, \citenamefont {Cavazzoni}, \citenamefont
  {Ceresoli}, \citenamefont {Chiarotti}, \citenamefont {Cococcioni},
  \citenamefont {Dabo}, \citenamefont {{Dal Corso}}, \citenamefont
  {de~Gironcoli}, \citenamefont {Fabris}, \citenamefont {Fratesi},
  \citenamefont {Gebauer}, \citenamefont {Gerstmann}, \citenamefont
  {Gougoussis}, \citenamefont {Kokalj}, \citenamefont {Lazzeri}, \citenamefont
  {Martin-Samos}, \citenamefont {Marzari}, \citenamefont {Mauri}, \citenamefont
  {Mazzarello}, \citenamefont {Paolini}, \citenamefont {Pasquarello},
  \citenamefont {Paulatto}, \citenamefont {Sbraccia}, \citenamefont {Scandolo},
  \citenamefont {Sclauzero}, \citenamefont {Seitsonen}, \citenamefont
  {Smogunov}, \citenamefont {Umari},\ and\ \citenamefont
  {Wentzcovitch}}]{QE2009}%
  \BibitemOpen
  \bibfield  {author} {\bibinfo {author} {\bibfnamefont {P.}~\bibnamefont
  {Giannozzi}}, \bibinfo {author} {\bibfnamefont {S.}~\bibnamefont {Baroni}},
  \bibinfo {author} {\bibfnamefont {N.}~\bibnamefont {Bonini}}, \bibinfo
  {author} {\bibfnamefont {M.}~\bibnamefont {Calandra}}, \bibinfo {author}
  {\bibfnamefont {R.}~\bibnamefont {Car}}, \bibinfo {author} {\bibfnamefont
  {C.}~\bibnamefont {Cavazzoni}}, \bibinfo {author} {\bibfnamefont
  {D.}~\bibnamefont {Ceresoli}}, \bibinfo {author} {\bibfnamefont {G.~L.}\
  \bibnamefont {Chiarotti}}, \bibinfo {author} {\bibfnamefont {M.}~\bibnamefont
  {Cococcioni}}, \bibinfo {author} {\bibfnamefont {I.}~\bibnamefont {Dabo}},
  \bibinfo {author} {\bibfnamefont {A.}~\bibnamefont {{Dal Corso}}}, \bibinfo
  {author} {\bibfnamefont {S.}~\bibnamefont {de~Gironcoli}}, \bibinfo {author}
  {\bibfnamefont {S.}~\bibnamefont {Fabris}}, \bibinfo {author} {\bibfnamefont
  {G.}~\bibnamefont {Fratesi}}, \bibinfo {author} {\bibfnamefont
  {R.}~\bibnamefont {Gebauer}}, \bibinfo {author} {\bibfnamefont
  {U.}~\bibnamefont {Gerstmann}}, \bibinfo {author} {\bibfnamefont
  {C.}~\bibnamefont {Gougoussis}}, \bibinfo {author} {\bibfnamefont
  {A.}~\bibnamefont {Kokalj}}, \bibinfo {author} {\bibfnamefont
  {M.}~\bibnamefont {Lazzeri}}, \bibinfo {author} {\bibfnamefont
  {L.}~\bibnamefont {Martin-Samos}}, \bibinfo {author} {\bibfnamefont
  {N.}~\bibnamefont {Marzari}}, \bibinfo {author} {\bibfnamefont
  {F.}~\bibnamefont {Mauri}}, \bibinfo {author} {\bibfnamefont
  {R.}~\bibnamefont {Mazzarello}}, \bibinfo {author} {\bibfnamefont
  {S.}~\bibnamefont {Paolini}}, \bibinfo {author} {\bibfnamefont
  {A.}~\bibnamefont {Pasquarello}}, \bibinfo {author} {\bibfnamefont
  {L.}~\bibnamefont {Paulatto}}, \bibinfo {author} {\bibfnamefont
  {C.}~\bibnamefont {Sbraccia}}, \bibinfo {author} {\bibfnamefont
  {S.}~\bibnamefont {Scandolo}}, \bibinfo {author} {\bibfnamefont
  {G.}~\bibnamefont {Sclauzero}}, \bibinfo {author} {\bibfnamefont {A.~P.}\
  \bibnamefont {Seitsonen}}, \bibinfo {author} {\bibfnamefont {A.}~\bibnamefont
  {Smogunov}}, \bibinfo {author} {\bibfnamefont {P.}~\bibnamefont {Umari}}, \
  and\ \bibinfo {author} {\bibfnamefont {R.~M.}\ \bibnamefont {Wentzcovitch}},\
  }\href {http://www.quantum-espresso.org} {\bibfield  {journal} {\bibinfo
  {journal} {Journal of Physics: Condensed Matter}\ }\textbf {\bibinfo {volume}
  {21}},\ \bibinfo {pages} {395502 (19pp)} (\bibinfo {year}
  {2009})}\BibitemShut {NoStop}%
\bibitem [{\citenamefont {Bl{\"o}chl}(1994)}]{Blochl1994}%
  \BibitemOpen
  \bibfield  {author} {\bibinfo {author} {\bibfnamefont {P.~E.}\ \bibnamefont
  {Bl{\"o}chl}},\ }\href {http://dx.doi.org/10.1103/PhysRevB.50.17953}
  {\bibfield  {journal} {\bibinfo  {journal} {Physical Review B}\ }\textbf
  {\bibinfo {volume} {50}},\ \bibinfo {pages} {17953} (\bibinfo {year}
  {1994})}\BibitemShut {NoStop}%
\bibitem [{\citenamefont {Heyd}\ \emph {et~al.}(2003)\citenamefont {Heyd},
  \citenamefont {Scuseria},\ and\ \citenamefont {Ernzerhof}}]{Heyd2003}%
  \BibitemOpen
  \bibfield  {author} {\bibinfo {author} {\bibfnamefont {J.}~\bibnamefont
  {Heyd}}, \bibinfo {author} {\bibfnamefont {G.~E.}\ \bibnamefont {Scuseria}},
  \ and\ \bibinfo {author} {\bibfnamefont {M.}~\bibnamefont {Ernzerhof}},\
  }\href {http://dx.doi.org/10.1063/1.1564060} {\bibfield  {journal} {\bibinfo
  {journal} {J. Chem. Phys.}\ }\textbf {\bibinfo {volume} {118}},\ \bibinfo
  {pages} {8207} (\bibinfo {year} {2003})}\BibitemShut {NoStop}%
\bibitem [{\citenamefont {Saraiva}\ \emph {et~al.}(2011)\citenamefont
  {Saraiva}, \citenamefont {Calder{\'o}n}, \citenamefont {Capaz}, \citenamefont
  {Hu}, \citenamefont {{Das Sarma}},\ and\ \citenamefont
  {Koiller}}]{Saraiva2011}%
  \BibitemOpen
  \bibfield  {author} {\bibinfo {author} {\bibfnamefont {A.~L.}\ \bibnamefont
  {Saraiva}}, \bibinfo {author} {\bibfnamefont {M.~J.}\ \bibnamefont
  {Calder{\'o}n}}, \bibinfo {author} {\bibfnamefont {R.~B.}\ \bibnamefont
  {Capaz}}, \bibinfo {author} {\bibfnamefont {X.}~\bibnamefont {Hu}}, \bibinfo
  {author} {\bibfnamefont {S.}~\bibnamefont {{Das Sarma}}}, \ and\ \bibinfo
  {author} {\bibfnamefont {B.}~\bibnamefont {Koiller}},\ }\href
  {http://dx.doi.org/10.1103/PhysRevB.84.155320} {\bibfield  {journal}
  {\bibinfo  {journal} {Physical Review B}\ }\textbf {\bibinfo {volume} {84}},\
  \bibinfo {pages} {155320} (\bibinfo {year} {2011})}\BibitemShut {NoStop}%
\bibitem [{\citenamefont {Heyd}\ \emph {et~al.}(2005)\citenamefont {Heyd},
  \citenamefont {Peralta}, \citenamefont {Scuseria},\ and\ \citenamefont
  {Martin}}]{Heyd2005}%
  \BibitemOpen
  \bibfield  {author} {\bibinfo {author} {\bibfnamefont {J.}~\bibnamefont
  {Heyd}}, \bibinfo {author} {\bibfnamefont {J.~E.}\ \bibnamefont {Peralta}},
  \bibinfo {author} {\bibfnamefont {G.~E.}\ \bibnamefont {Scuseria}}, \ and\
  \bibinfo {author} {\bibfnamefont {R.~L.}\ \bibnamefont {Martin}},\ }\href
  {http://dx.doi.org/10.1063/1.2085170} {\bibfield  {journal} {\bibinfo
  {journal} {J. Chem. Phys.}\ }\textbf {\bibinfo {volume} {123}},\ \bibinfo
  {pages} {174101} (\bibinfo {year} {2005})}\BibitemShut {NoStop}%
\bibitem [{\citenamefont {Kohn}\ and\ \citenamefont {Sham}(1965)}]{Kohn1965}%
  \BibitemOpen
  \bibfield  {author} {\bibinfo {author} {\bibfnamefont {W.}~\bibnamefont
  {Kohn}}\ and\ \bibinfo {author} {\bibfnamefont {L.~J.}\ \bibnamefont
  {Sham}},\ }\href {http://dx.doi.org/10.1103/PhysRev.140.A1133} {\bibfield
  {journal} {\bibinfo  {journal} {Physical Review}\ }\textbf {\bibinfo {volume}
  {140}},\ \bibinfo {pages} {A1133} (\bibinfo {year} {1965})}\BibitemShut
  {NoStop}%
\bibitem [{\citenamefont {Castner}(2009)}]{Castner2009}%
  \BibitemOpen
  \bibfield  {author} {\bibinfo {author} {\bibfnamefont {T.~G.}\ \bibnamefont
  {Castner}},\ }\href {\doibase 10.1103/PhysRevB.79.195207} {\bibfield
  {journal} {\bibinfo  {journal} {Phys. Rev. B}\ }\textbf {\bibinfo {volume}
  {79}},\ \bibinfo {pages} {195207} (\bibinfo {year} {2009})}\BibitemShut
  {NoStop}%
\bibitem [{\citenamefont {Greenman}\ \emph {et~al.}(2013)\citenamefont
  {Greenman}, \citenamefont {Whitley},\ and\ \citenamefont
  {Whaley}}]{Greenman2013}%
  \BibitemOpen
  \bibfield  {author} {\bibinfo {author} {\bibfnamefont {L.}~\bibnamefont
  {Greenman}}, \bibinfo {author} {\bibfnamefont {H.~D.}\ \bibnamefont
  {Whitley}}, \ and\ \bibinfo {author} {\bibfnamefont {K.~B.}\ \bibnamefont
  {Whaley}},\ }\href {\doibase 10.1103/PhysRevB.88.165102} {\bibfield
  {journal} {\bibinfo  {journal} {Phys. Rev. B}\ }\textbf {\bibinfo {volume}
  {88}},\ \bibinfo {pages} {165102} (\bibinfo {year} {2013})}\BibitemShut
  {NoStop}%
\bibitem [{\citenamefont {Fritzsche}(1962)}]{Fritzsche1962}%
  \BibitemOpen
  \bibfield  {author} {\bibinfo {author} {\bibfnamefont {H.}~\bibnamefont
  {Fritzsche}},\ }\href {\doibase 10.1103/PhysRev.125.1560} {\bibfield
  {journal} {\bibinfo  {journal} {Phys. Rev.}\ }\textbf {\bibinfo {volume}
  {125}},\ \bibinfo {pages} {1560} (\bibinfo {year} {1962})}\BibitemShut
  {NoStop}%
\bibitem [{\citenamefont {Baena}\ \emph {et~al.}(2012)\citenamefont {Baena},
  \citenamefont {Saraiva}, \citenamefont {Koiller},\ and\ \citenamefont
  {Calder\'{o}n}}]{Baena2012}%
  \BibitemOpen
  \bibfield  {author} {\bibinfo {author} {\bibfnamefont {A.}~\bibnamefont
  {Baena}}, \bibinfo {author} {\bibfnamefont {A.~L.}\ \bibnamefont {Saraiva}},
  \bibinfo {author} {\bibfnamefont {B.}~\bibnamefont {Koiller}}, \ and\
  \bibinfo {author} {\bibfnamefont {M.~J.}\ \bibnamefont {Calder\'{o}n}},\
  }\href {\doibase 10.1103/PhysRevB.86.035317} {\bibfield  {journal} {\bibinfo
  {journal} {Phys. Rev. B}\ }\textbf {\bibinfo {volume} {86}},\ \bibinfo
  {pages} {035317} (\bibinfo {year} {2012})}\BibitemShut {NoStop}%
\bibitem [{\citenamefont {Hu}\ \emph {et~al.}(2005)\citenamefont {Hu},
  \citenamefont {Koiller},\ and\ \citenamefont {Das~Sarma}}]{Hu2005}%
  \BibitemOpen
  \bibfield  {author} {\bibinfo {author} {\bibfnamefont {X.}~\bibnamefont
  {Hu}}, \bibinfo {author} {\bibfnamefont {B.}~\bibnamefont {Koiller}}, \ and\
  \bibinfo {author} {\bibfnamefont {S.}~\bibnamefont {Das~Sarma}},\ }\href
  {\doibase 10.1103/PhysRevB.71.235332} {\bibfield  {journal} {\bibinfo
  {journal} {Phys. Rev. B}\ }\textbf {\bibinfo {volume} {71}},\ \bibinfo
  {pages} {235332} (\bibinfo {year} {2005})}\BibitemShut {NoStop}%
\bibitem [{\citenamefont {Koiller}\ \emph {et~al.}(2001)\citenamefont
  {Koiller}, \citenamefont {Hu},\ and\ \citenamefont
  {Das~Sarma}}]{Koiller2001}%
  \BibitemOpen
  \bibfield  {author} {\bibinfo {author} {\bibfnamefont {B.}~\bibnamefont
  {Koiller}}, \bibinfo {author} {\bibfnamefont {X.}~\bibnamefont {Hu}}, \ and\
  \bibinfo {author} {\bibfnamefont {S.}~\bibnamefont {Das~Sarma}},\ }\href
  {\doibase 10.1103/PhysRevLett.88.027903} {\bibfield  {journal} {\bibinfo
  {journal} {Phys. Rev. Lett.}\ }\textbf {\bibinfo {volume} {88}},\ \bibinfo
  {pages} {027903} (\bibinfo {year} {2001})}\BibitemShut {NoStop}%
\bibitem [{\citenamefont {Li}\ \emph {et~al.}(2010)\citenamefont {Li},
  \citenamefont {Cywi\ifmmode~\acute{n}\else \'{n}\fi{}ski}, \citenamefont
  {Culcer}, \citenamefont {Hu},\ and\ \citenamefont {Das~Sarma}}]{Li2010}%
  \BibitemOpen
  \bibfield  {author} {\bibinfo {author} {\bibfnamefont {Q.}~\bibnamefont
  {Li}}, \bibinfo {author} {\bibfnamefont {L.}~\bibnamefont
  {Cywi\ifmmode~\acute{n}\else \'{n}\fi{}ski}}, \bibinfo {author}
  {\bibfnamefont {D.}~\bibnamefont {Culcer}}, \bibinfo {author} {\bibfnamefont
  {X.}~\bibnamefont {Hu}}, \ and\ \bibinfo {author} {\bibfnamefont
  {S.}~\bibnamefont {Das~Sarma}},\ }\href {\doibase 10.1103/PhysRevB.81.085313}
  {\bibfield  {journal} {\bibinfo  {journal} {Phys. Rev. B}\ }\textbf {\bibinfo
  {volume} {81}},\ \bibinfo {pages} {085313} (\bibinfo {year}
  {2010})}\BibitemShut {NoStop}%
\bibitem [{\citenamefont {Oberbeck}\ \emph {et~al.}(2004)\citenamefont
  {Oberbeck}, \citenamefont {Curson}, \citenamefont {Hallam}, \citenamefont
  {Simmons}, \citenamefont {Bilger},\ and\ \citenamefont
  {Clark}}]{Oberbeck2004}%
  \BibitemOpen
  \bibfield  {author} {\bibinfo {author} {\bibfnamefont {L.}~\bibnamefont
  {Oberbeck}}, \bibinfo {author} {\bibfnamefont {N.~J.}\ \bibnamefont
  {Curson}}, \bibinfo {author} {\bibfnamefont {T.}~\bibnamefont {Hallam}},
  \bibinfo {author} {\bibfnamefont {M.~Y.}\ \bibnamefont {Simmons}}, \bibinfo
  {author} {\bibfnamefont {G.}~\bibnamefont {Bilger}}, \ and\ \bibinfo {author}
  {\bibfnamefont {R.~G.}\ \bibnamefont {Clark}},\ }\href {\doibase
  10.1063/1.1784881} {\bibfield  {journal} {\bibinfo  {journal} {Appl. Phys.
  Lett.}\ }\textbf {\bibinfo {volume} {85}},\ \bibinfo {pages} {1359} (\bibinfo
  {year} {2004})}\BibitemShut {NoStop}%
\bibitem [{\citenamefont {Bielejec}\ \emph {et~al.}(2010)\citenamefont
  {Bielejec}, \citenamefont {Seamons},\ and\ \citenamefont
  {Carroll}}]{Bielejec2010}%
  \BibitemOpen
  \bibfield  {author} {\bibinfo {author} {\bibfnamefont {E.}~\bibnamefont
  {Bielejec}}, \bibinfo {author} {\bibfnamefont {J.~A.}\ \bibnamefont
  {Seamons}}, \ and\ \bibinfo {author} {\bibfnamefont {M.~S.}\ \bibnamefont
  {Carroll}},\ }\href {http://stacks.iop.org/0957-4484/21/i=8/a=085201}
  {\bibfield  {journal} {\bibinfo  {journal} {Nanotechnology}\ }\textbf
  {\bibinfo {volume} {21}},\ \bibinfo {pages} {085201} (\bibinfo {year}
  {2010})}\BibitemShut {NoStop}%
\bibitem [{\citenamefont {Boykin}\ \emph {et~al.}(2004)\citenamefont {Boykin},
  \citenamefont {Klimeck},\ and\ \citenamefont {Oyafuso}}]{Boykin2004}%
  \BibitemOpen
  \bibfield  {author} {\bibinfo {author} {\bibfnamefont {T.~B.}\ \bibnamefont
  {Boykin}}, \bibinfo {author} {\bibfnamefont {G.}~\bibnamefont {Klimeck}}, \
  and\ \bibinfo {author} {\bibfnamefont {F.}~\bibnamefont {Oyafuso}},\ }\href
  {\doibase 10.1103/PhysRevB.69.115201} {\bibfield  {journal} {\bibinfo
  {journal} {Phys. Rev. B}\ }\textbf {\bibinfo {volume} {69}},\ \bibinfo
  {pages} {115201} (\bibinfo {year} {2004})}\BibitemShut {NoStop}%
\bibitem [{\citenamefont {Rahman}\ \emph {et~al.}(2011)\citenamefont {Rahman},
  \citenamefont {Park}, \citenamefont {Klimeck},\ and\ \citenamefont
  {Hollenberg}}]{Rahman2011}%
  \BibitemOpen
  \bibfield  {author} {\bibinfo {author} {\bibfnamefont {R.}~\bibnamefont
  {Rahman}}, \bibinfo {author} {\bibfnamefont {S.~H.}\ \bibnamefont {Park}},
  \bibinfo {author} {\bibfnamefont {G.}~\bibnamefont {Klimeck}}, \ and\
  \bibinfo {author} {\bibfnamefont {L.~C.~L.}\ \bibnamefont {Hollenberg}},\
  }\href {\doibase 10.1088/0957-4484/22/22/225202} {\bibfield  {journal}
  {\bibinfo  {journal} {Nanotechnology}\ }\textbf {\bibinfo {volume} {22}},\
  \bibinfo {pages} {225202} (\bibinfo {year} {2011})}\BibitemShut {NoStop}%
\end{thebibliography}%

\end{document}